\documentclass[11pt]{article}
\pdfoutput=1

\usepackage[margin=1in, paperwidth=8.5in, paperheight=11in]{geometry}

\usepackage{graphicx}
\usepackage{float}
\usepackage{caption}
\usepackage{subcaption}
\usepackage{fp}
\usepackage{pgfplots}
\pgfplotsset{compat=1.13}

\usepackage{tikz}

\usepackage{verbatim}

\usepackage{booktabs}

\usepackage[numbers]{natbib}
\usepackage{url}

\begin{document}

\title{ Applying Multi-qubit Correction to Frustrated Cluster Loops on an Adiabatic Quantum Computer }
\author{John E. Dorband\\
	Department of Computer Science and Electrical Engineering\\
	University of Maryland, Baltimore County\\
	Maryland, USA\\
	\texttt{dorband@umbc.edu}}
\date{\today}
\maketitle

\begin{abstract}

The class of problems represented by frustrated cluster loops, FCL, is a robust set of problems that spans a wide
range of computational difficulty and that are easy to determine what their solutions are. 
Here, we use frustrated cluster loops to test the relative performance of the D-Wave without post-processing
and the D-Wave with multi-qubit correction (MQC) post-processing.
MQC post-processing has shown itself exceptionally beneficial in improving the performance of the D-Wave 2000Q
when processing difficult FCL problems.

\end{abstract}

\section{Introduction}\label{sec:intro}

The D-Wave\citep{Dwave13} is an adiabatic quantum computer\citep{Farhi00,Giuseppe08} which supports the
following objective function:
\begin{equation}\label{eq:obfunc}
F = {\sum\limits_i a_i q_i + \sum\limits_i \sum\limits_j b_{ij} q_i q_j}
\end{equation}
where $q_i\in\{-1,1\}$ are the qubit values returned by the D-Wave, and $a_i\in[-2,2]$ and $b_{ij}\in[-1,1]$ are 
the coefficients given to the D-Wave associated with the qubits and the qubit couplers respectively. 

An algorithm has been developed, multi-qubit correction, MQC, that
reduces a set of D-Wave samples to a single sample that has an objective function value that is less-than
or equal-to the objective function value of any sample in the D-Wave sample set.
This algorithm presumes that the D-Wave samples contain groups of qubits that can be used to construct
a more optimal solution to the objective function. 
This has borne out to be true, as has been presented in the paper \citep{Dorband17}.

The problem class designated as frustrated cluster loops have been used to show how well a quantum annealer's
performance compares to that of a simulated annealer running on a classical computer.
Two such methods are described in \citep{Hen2015} and \citep{King2015}.
Other similar problems have been constructed to demonstrate quantum speedup such as \citep{King2017} 
and \citep{Mandra18}, but will not be utilized here.  
These problems have been developed to show if and when a quantum annealer such as the D-Wave 
exceeds the performance of a classical simulated annealer.

The purpose here however is not to show how the D-Wave compares to classical computers, but to show how the D-wave
performs with and without MQC post-processing.
The two problem classes that will be used in the comparison will be designated as 
type 1 described in \citep{Hen2015} and type 2 described in \citep{King2015}.  
Granted this is not aimed at showing quantum supremacy, but is to show how much faster the optimal
result can be obtained on the D-Wave with MQC post-processing than with the D-Wave alone.

The advantage of using frustrated cluster loops (FCL) as a means of testing an adiabatic quantum computer is 
two fold, 1) the solution to an FCL problem is easily determined before it is run and 2) cases of FCL can be easily
generated that have wide variations in their solution difficulty.
The solution to an FCL problem (the global minimum of the objective function) is simply the sum of all the 
qubit and coupler coefficients.  
The difficulty of an FCL problem is measured by how many samples it takes for it to be solved. 
Figures \ref{fig:STSRr} and \ref{fig:STSRCr} show plots of the $log_2$ of how many samples it takes for the
D-Wave to solve various cases of FCL problems.
They range from 1 sample to $2^{13} = 8192$ samples.
And in some cases the D-Wave never found the solution.

\section{The Frustrated Cluster Loop Problem}\label{sec:fcl}

A frustrated cluster loop (FCL) problem on the D-Wave consists of creating randomly generated loops 
over a group of qubits.
A loop consists of a chain of qubits interleaved with couplers connected in a closed loop.
These loops take random paths of random length around the group of qubits.
The qubit coefficients are given a value of zero and the coupler coefficients are given a value of -1, except
one random coupler which is given a coupler coefficient value of 1.
A number of loops are generated for the group of qubits equal to the number of qubits in the group
times a constant, $\alpha$.
The qubit and coupler coefficients of the loops are summed together for each corresponding qubit and coupler in
the group.
This forms the problem to be optimized on the D-Wave.
For the D-Wave, the groups of qubits are specified in terms of square regions of qubit cells.
A region of size $L_c$ designates a qubit group of size $c$ by $c$ cells of 8 qubits each.
Thus for a qubit group of size $L_4$, 128 qubits, and an  $\alpha$ value of 0.1, 12 or 13 loops would be randomly
generated for a single frustrated cluster loop problem.

The difference between a type 1 problem and a type 2 problem is how a minimum size problem is determined.
A type 1 problem must contain at least 8 qubits, while for a type 2 problem a loop must contain qubits from more 
than one cell.
This allows a type 2 problem to contain as few as 6 qubits.

An overlap factor, $R$, can also be specified.
The overlap factor indicates the maximum number of loops that can include a specific coupler.
If, as loops are being generated, a loop attempts to use a coupler that already is being used by $R$ loops, 
that loop must be discarded.
$R$ is a metric of ruggedness.
The larger $R$ is, the harder it is for hardware (D-Wave) to represent the problem coefficients accurately.
The values of $R$ used here are 2, 3, and $\infty$ (unlimited).

\section{Testing MQC With Frustrated Cluster Loops }\label{sec:method}

The frustrated cluster loop problem is used here to compare the performance of the D-Wave to the D-Wave with
MQC post-processing.
MQC takes a group of D-Wave samples and reduces them to a single sample as described in \citep{Dorband17}.
Since each sample takes a fixed time, about $20\mu s$, the number of samples needed to solve a problem
will be used in place of a metric of time.

For the D-Wave without MQC post-processing, a solution is determined to be found within a number of samples,
if one of those samples is a solution.
For the D-Wave with MQC post-processing, a solution is determined to be found within a number of samples,
if those samples have been reduced to a solution.

\subsection{Initial Test}\label{sec:initTest}

\begin{figure}
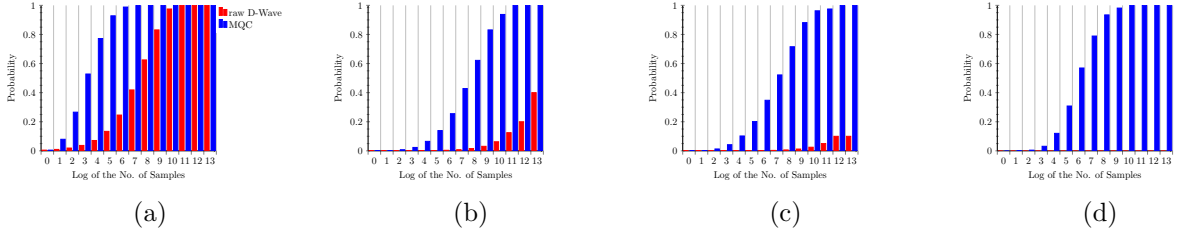

    \begin{subfigure}{0.24\textwidth}
	\scalebox{0.34}{
	    \begin{tikzpicture}
		\input{tab/accumulated}
		  \begin{axis} [
      axis lines=left,
      xlabel={Log of the No. of Samples},
      ylabel={Probability},
      ymin=0, ymax=1.0,
      xmin=0, xmax=14,
      minor y tick num=3,
      minor x tick num=0,
      legend cell align=left,
      legend style={ draw=none, at={(1.40,1.0)}, },
      ybar interval=0.7,
    ]
    \addplot[red,fill=red]     table[ x index=0, y index=2, ]                {\plotdata};
    \addplot[blue,fill=blue]   table[ x index=0, y index=3, ]                {\plotdata};
    \legend{raw D-Wave,MQC}
  \end{axis}
	    \end{tikzpicture}
	}
	\caption{  } \label{fig:Case_1_0}
    \end{subfigure}
    \hspace*{\fill} 
    \begin{subfigure}{0.23\textwidth}
	\scalebox{0.34}{
	    \begin{tikzpicture}
		\input{tab/accumulated}
		  \begin{axis} [
      axis lines=left,
      xlabel={Log of the No. of Samples},
      ylabel={Probability},
      ymin=0, ymax=1.0,
      xmin=0, xmax=14,
      minor y tick num=3,
      minor x tick num=0,
      legend cell align=left,
      legend style={ draw=none, at={(1.35,1.0)}, },
      ybar interval=0.7,
    ]
    \addplot[red,fill=red]     table[ x index=0, y index=10, ]                {\plotdata};
    \addplot[blue,fill=blue]   table[ x index=0, y index=11, ]                {\plotdata};
  \end{axis}
	    \end{tikzpicture}
	}
	\caption{  } \label{fig:Case_1_4}
    \end{subfigure}
    \hspace*{\fill} 
    \begin{subfigure}{0.23\textwidth}
	\scalebox{0.34}{
	    \begin{tikzpicture}
		\input{tab/accumulated}
		  \begin{axis} [
      axis lines=left,
      xlabel={Log of the No. of Samples},
      ylabel={Probability},
      ymin=0, ymax=1.0,
      xmin=0, xmax=14,
      minor y tick num=3,
      minor x tick num=0,
      legend cell align=left,
      legend style={ draw=none, at={(1.35,1.0)}, },
      ybar interval=0.7,
    ]
    \addplot[red,fill=red]     table[ x index=0, y index=12, ]                {\plotdata};
    \addplot[blue,fill=blue]   table[ x index=0, y index=13, ]                {\plotdata};
  \end{axis}
	    \end{tikzpicture}
	}
	\caption{  } \label{fig:Case_1_5}
    \end{subfigure}
    \hspace*{\fill} 
    \begin{subfigure}{0.23\textwidth}
	\scalebox{0.34}{
	    \begin{tikzpicture}
		\input{tab/accumulated}
		  \begin{axis} [
      axis lines=left,
      xlabel={Log of the No. of Samples},
      ylabel={Probability},
      ymin=0, ymax=1.0,
      xmin=0, xmax=14,
      minor y tick num=3,
      minor x tick num=0,
      legend cell align=left,
      legend style={ draw=none, at={(1.35,1.0)}, },
      ybar interval=0.7,
    ]
    \addplot[red,fill=red]     table[ x index=0, y index=8, ]                {\plotdata};
    \addplot[blue,fill=blue]   table[ x index=0, y index=9, ]                {\plotdata};
  \end{axis}
	    \end{tikzpicture}
	}
	\caption{  } \label{fig:Case_1_3}
    \end{subfigure}
     \caption{ Comparison of the performance of D-Wave verses D-Wave with MQC post-processing for
	several case where $\alpha=0.1$. Each case represents 81920 samples. 
	The number samples to solution are displayed as the log of the number of samples. }
    \label{fig:Cases0}
\end{figure}

\begin{figure}
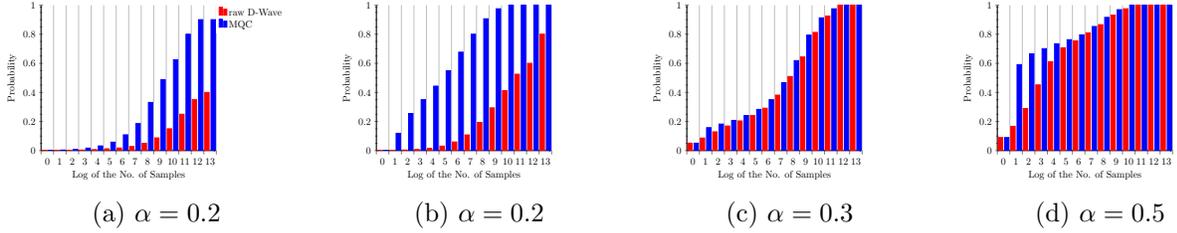

    \begin{subfigure}{0.25\textwidth}
	\scalebox{0.34}{
	    \begin{tikzpicture}
		\input{tab/accumulated}
		  \begin{axis} [
      axis lines=left,
      xlabel={Log of the No. of Samples},
      ylabel={Probability},
      ymin=0, ymax=1.0,
      xmin=0, xmax=14,
      minor y tick num=3,
      minor x tick num=0,
      legend cell align=left,
      legend style={ draw=none, at={(1.40,1.0)}, },
      ybar interval=0.7,
    ]
    \addplot[red,fill=red]     table[ x index=0, y index=34, ]                {\plotdata};
    \addplot[blue,fill=blue]   table[ x index=0, y index=35, ]                {\plotdata};
    \legend{raw D-Wave,MQC}
  \end{axis}
	    \end{tikzpicture}
	}
	\caption{ $\alpha=0.2$ } \label{fig:Case_2_6}
    \end{subfigure}
    \hspace*{\fill} 
    \begin{subfigure}{0.23\textwidth}
	\scalebox{0.34}{
	    \begin{tikzpicture}
		\input{tab/accumulated}
		  \begin{axis} [
      axis lines=left,
      xlabel={Log of the No. of Samples},
      ylabel={Probability},
      ymin=0, ymax=1.0,
      xmin=0, xmax=14,
      minor y tick num=3,
      minor x tick num=0,
      legend cell align=left,
      legend style={ draw=none, at={(1.35,1.0)}, },
      ybar interval=0.7,
    ]
    \addplot[red,fill=red]     table[ x index=0, y index=40, ]                {\plotdata};
    \addplot[blue,fill=blue]   table[ x index=0, y index=41, ]                {\plotdata};
  \end{axis}
	    \end{tikzpicture}
	}
	\caption{ $\alpha=0.2$ } \label{fig:Case_2_9}
    \end{subfigure}
    \hspace*{\fill} 
    \begin{subfigure}{0.23\textwidth}
	\scalebox{0.34}{
	    \begin{tikzpicture}
		\input{tab/accumulated}
		  \begin{axis} [
      axis lines=left,
      xlabel={Log of the No. of Samples},
      ylabel={Probability},
      ymin=0, ymax=1.0,
      xmin=0, xmax=14,
      minor y tick num=3,
      minor x tick num=0,
      legend cell align=left,
      legend style={ draw=none, at={(1.35,1.0)}, },
      ybar interval=0.7,
    ]
    \addplot[red,fill=red]     table[ x index=0, y index=54, ]                {\plotdata};
    \addplot[blue,fill=blue]   table[ x index=0, y index=55, ]                {\plotdata};
  \end{axis}
	    \end{tikzpicture}
	}
	\caption{ $\alpha=0.3$ } \label{fig:Case_3_6}
    \end{subfigure}
    \hspace*{\fill} 
    \begin{subfigure}{0.23\textwidth}
	\scalebox{0.34}{
	    \begin{tikzpicture}
		\input{tab/accumulated}
		  \begin{axis} [
      axis lines=left,
      xlabel={Log of the No. of Samples},
      ylabel={Probability},
      ymin=0, ymax=1.0,
      xmin=0, xmax=14,
      minor y tick num=3,
      minor x tick num=0,
      legend cell align=left,
      legend style={ draw=none, at={(1.35,1.0)}, },
      ybar interval=0.7,
    ]
    \addplot[red,fill=red]     table[ x index=0, y index=84, ]                {\plotdata};
    \addplot[blue,fill=blue]   table[ x index=0, y index=85, ]                {\plotdata};
  \end{axis}
	    \end{tikzpicture}
	}
	\caption{ $\alpha=0.5$ } \label{fig:Case_5_1}
    \end{subfigure}
     \caption{ Comparison of the performance of D-Wave alone and D-Wave with MQC 
     post-processing where each case represents 81920 samples.
     The number samples are displayed as the log of the number of samples. }
    \label{fig:Cases1}
\end{figure}

The purpose of this initial test is to show the diversity in the complexity and difficulty of the frustrated 
cluster loop class of problems. 
The difficulty of a case of this problem class is demonstrated by how quickly the D-Wave solves it or 
whether it solves it at all.
Figure \ref{fig:Case_3_6} is an example of a case that the D-Wave solves easily,
while figure \ref{fig:Case_1_3} is an example of a hard case that the D-Wave did not solve within 81920 samples.
The initial test was performed for the following problem generation parameters, $R=\infty$, $c=16$, 
and for $\alpha$ between 0.1 to 0.5 in increments of 0.1.
50 cases were randomly generated, 10 cases for each of the 5 values of $\alpha$.
81920 samples were requested from a 2048 qubit D-Wave 2000Q for each case.

Figures  \ref{fig:Cases0} and  \ref{fig:Cases1} show arbtrary examples of FCL problems that demonstrate the 
diversity of relative behavior of the D-Wave with MQC post-processing and without MQC post-processing.
All the cases in \ref{fig:Cases0} where created with a value of $\alpha=0.1$.
In each of these case a probability of 1.0 indicates that for every group of size $2^n$ samples, a sample was 
attained with an objective function value equal to the global minimum.
A probability of 0.5 indicates that only half of the groups of size $2^n$ samples attained the global minimum.
In figure \ref{fig:Case_1_0}, the D-Wave attained the global minimum consistently for sample groups larger than
$2^{11}=2048$ samples while the D-Wave with MQC post-processing attained the global minimum consistently for
sample groups of size $2^7=128$ or larger. 
Therefore MQC post-processing increased the speed by a factor of 16.
In figure \ref{fig:Case_1_3}, out of 81920 samples the D-Wave never found the global minimum, while MQC 
consistently found the global minimum with 1024 samples. 
Figure \ref{fig:Cases0} shows a range of difficulty of cases, 
from \ref{fig:Case_1_0} which is moderately hard to \ref{fig:Case_1_3} which was very hard.
Figure \ref{fig:Cases1} shows cases with different values of $\alpha$ where \ref{fig:Case_3_6} 
and \ref{fig:Case_5_1} are very easy and gain very little benefit from the use of MQC 
and \ref{fig:Case_2_6} and  \ref{fig:Case_2_9} which are much harder to the extent that MQC finds the global 
minimum for \ref{fig:Case_2_6} only 90\% of the time with a group size of 8192 samples and the D-Wave alone
only 40\% of the time.

\subsection{Problem Class Survey}\label{sec:svyTest}

In section \ref{sec:initTest}, for each case, 81920 samples were obtained from the D-Wave for analysis.
This section surveys a much larger parameter space so only 8192 samples were obtained for each case.
The survey is performed over various ranges of the parameters $R$, $L_c$, and $\alpha$.
The values of $R$ are 2, 3, and unlimited ($\infty$), 
the values of $c$ for $L_c$ are 2, 3, 4, ... 16, and
the values of $\alpha$ range from 0.05 to 0.5 in increments of 0.05.
There were 100 cases generated for each $R$, $L_c$, and $\alpha$.
For each case, 8192 samples were requested from a 2048 qubit D-Wave 2000Q.

For each case the 8192 samples were divided into groups, 8192 groups of size 1, 4096 groups of size 2,
2048 groups of size 4, and so forth up to 2 groups of size 4096 and 1 group of size 8192, in powers of two.
Each group of D-Wave samples is consider to have succeeded if at least one sample in the group has an
objective function value that is equal to the global minimum.
Each group of D-Wave samples, post-processed with MQC, is considered to have succeeded if its resultant 
sample has an objective function value that is equal to the global minimum.

\begin{figure}
    \begin{subfigure}{0.32\textwidth}
	\scalebox{0.48}{
	    \begin{tikzpicture}
		  \pgfplotstableread{
 Alpha       rawL02       mqcL02       rawL03       mqcL03       rawL04       mqcL04       rawL05       mqcL05       rawL06       mqcL06       rawL07       mqcL07       rawL08       mqcL08       rawL09       mqcL09       rawL10       mqcL10       rawL11       mqcL11       rawL12       mqcL12       rawL13       mqcL13       rawL14       mqcL14       rawL15       mqcL15       rawL16       mqcL16
  0.05     0.097611     0.097611     1.691534     1.580145     2.381283     2.217231     2.922198     2.575312     3.235727     3.046142     3.568032     3.364572     4.310340     4.044394     4.627607     4.220330     5.072106     4.578939     5.617063     4.957915     6.272023     5.212569     6.560104     5.407353     6.932628     5.454176     7.349790     5.577731     7.500802     5.275007 
  0.10     1.327687     1.280956     2.709291     2.632268     3.056584     3.007196     4.215679     4.139142     4.440952     4.229588     5.416840     4.944858     6.455492     5.740928     6.771886     5.737687     7.615887     6.364572     8.444932     6.929791     8.859970     6.817623     9.358988     7.165108     9.942045     7.226509    11.057144     7.803227    11.820326     7.788686 
  0.15     1.443607     1.389567     2.622930     2.604071     3.214125     3.195348     4.234195     4.116032     4.684819     4.553361     5.263034     4.980025     6.422233     6.035624     7.118526     6.450221     7.584963     6.851999     7.822730     7.082362     8.221877     7.140779     9.090853     7.401903     9.932886     7.974529    10.375039     8.232661    11.246345     8.596935 
  0.20     1.333424     1.298658     2.084064     2.021480     2.669027     2.632268     3.613532     3.596935     4.438293     4.275007     4.906891     4.713696     5.743084     5.654206     6.283922     5.967169     6.577731     6.257011     6.555816     6.280956     7.424922     6.883621     7.283922     6.454176     8.274262     6.916477     8.399171     7.526069     7.996389     7.157044 
  0.25     1.195348     1.169925     1.613532     1.555816     1.722466     1.632268     2.731183     2.726831     3.528571     3.526069     4.012569     3.942984     5.368070     5.358256     5.681449     5.617063     5.348374     5.302319     4.928844     4.886550     5.957915     5.809414     6.702658     5.773996     7.489286     6.513491     7.412782     7.019702     6.378512     6.174726 
  0.30     1.117695     1.090853     1.275007     1.263034     1.432959     1.389567     1.855990     1.847997     2.731183     2.718088     3.277985     3.266037     4.147307     4.129283     4.733354     4.680324     4.812498     4.795975     3.925999     3.912650     4.893362     4.720278     6.063503     5.201634     5.823749     4.887525     6.742006     6.574102     5.625270     5.589763 
  0.35     0.839960     0.839960     1.007196     0.992768     1.084064     1.042644     1.244887     1.244887     1.790772     1.782409     2.077243     2.063503     2.182692     2.157044     3.650765     3.641546     3.319040     3.319040     3.060047     3.014355     3.669027     3.310340     5.526069     4.632268     5.333424     4.310340     5.653060     5.607626     4.427606     4.393691 
  0.40     0.887525     0.871844     0.855990     0.855990     0.963474     0.963474     1.104337     1.104337     1.350497     1.339137     1.655352     1.655352     1.454176     1.422233     3.080658     3.060047     3.021480     3.003602     2.405992     2.344828     3.480265     3.182692     5.443607     4.495695     5.321928     4.250962     4.811471     4.648465     4.051372     4.047887 
  0.45     0.799087     0.799087     0.863938     0.863938     0.925999     0.925999     1.014355     1.014355     1.292782     1.292782     1.773996     1.765535     1.505891     1.485427     3.049631     3.049631     2.565597     2.541019     2.207893     2.124328     3.405992     3.117695     5.505891     4.604071     5.201634     4.275007     5.063503     4.933573     3.735522     3.733354 
  0.50     0.790772     0.790772     0.823749     0.823749     0.948601     0.948601     1.104337     1.104337     1.056584     1.042644     1.739848     1.731183     1.411426     1.367371     3.395063     3.389567     2.286881     2.286881     2.250962     2.189034     3.232661     2.948601     5.226509     4.137504     5.163499     4.214125     5.304511     5.264536     3.881665     3.879706 
  }\plotdata;
		  \begin{axis} [
      axis lines=left,
      xlabel={$\alpha$},
      ylabel={$Log_2$ of samples},
      ymin=0, ymax=15,
      xmin=0.05, xmax=0.5,
      minor x tick num=1,
      minor y tick num=4,
      legend cell align=left,
      legend style={ draw=none, at={(1.35,1.0)}, },
    ]
    \addplot[red]    table[ x index=0, y index=1,  ]                {\plotdata};
    \addplot[cyan]   table[ x index=0, y index=5,  ]                {\plotdata};
    \addplot[violet] table[ x index=0, y index=9,  ]                {\plotdata};
    \addplot[teal]   table[ x index=0, y index=13, ]                {\plotdata};
    \addplot[blue]   table[ x index=0, y index=17, ]                {\plotdata};
    \addplot[orange] table[ x index=0, y index=21, ]                {\plotdata};
    \addplot[yellow] table[ x index=0, y index=25, ]                {\plotdata};
    \addplot[magenta]table[ x index=0, y index=29, ]                {\plotdata};
    \legend{$c = 2$,$c = 4$,$c = 6$,$c = 8$,$c = 10$,$c = 12$,$c = 14$,$c = 16$}
  \end{axis}
	    \end{tikzpicture}
	}
	\caption{ $R=2$ } \label{fig:STSRM02r}
    \end{subfigure}
    \hspace*{\fill} 
    \begin{subfigure}{0.32\textwidth}
	\scalebox{0.48}{
	    \begin{tikzpicture}
		  \pgfplotstableread{
 Alpha       rawL02       mqcL02       rawL03       mqcL03       rawL04       mqcL04       rawL05       mqcL05       rawL06       mqcL06       rawL07       mqcL07       rawL08       mqcL08       rawL09       mqcL09       rawL10       mqcL10       rawL11       mqcL11       rawL12       mqcL12       rawL13       mqcL13       rawL14       mqcL14       rawL15       mqcL15       rawL16       mqcL16
  0.05     0.226509     0.226509     1.613532     1.454176     2.601697     2.503349     3.021480     2.823749     3.117695     2.952334     4.258519     3.748461     4.853996     4.371559     4.746313     4.286881     5.812498     5.102658     6.160275     5.104337     6.607626     5.344828     7.017922     5.153805     7.503349     5.582556     7.747924     5.295723     9.015248     5.516015 
  0.10     1.480265     1.405992     2.669027     2.565597     3.744161     3.729009     4.189034     4.025029     5.082362     4.650765     5.713696     5.152183     6.798051     6.070389     7.142413     5.797013     8.147307     6.599318     9.393004     7.049631     9.589763     7.090853    10.580145     7.456806    10.799975     7.475085    12.208947     7.773996    12.351763     7.839960 
  0.15     1.752749     1.655352     2.678072     2.604071     3.897240     3.873813     4.589763     4.440952     5.324811     5.046142     5.790772     5.432959     7.100978     6.411426     7.363171     6.341986     8.203201     7.137504     9.291309     7.498251     9.710493     7.843984    10.171527     7.851999    10.818760     8.003602    12.208995     8.761285    12.051225     8.615887 
  0.20     1.669027     1.604071     2.469886     2.356144     3.324811     3.263034     4.134221     4.054848     4.952334     4.778209     5.558268     5.316146     6.364572     6.028569     6.925999     6.250962     6.994580     6.505891     7.967169     7.275007     8.552131     7.357552     9.099295     7.587365     9.638074     7.648465    10.941106     8.451541    10.918620     8.422233 
  0.25     1.641546     1.575312     2.130931     2.007196     3.147307     3.056584     3.726831     3.698218     4.260026     4.155425     5.037382     4.942984     5.829850     5.698218     6.087463     5.678072     6.657640     6.330558     7.241840     6.893362     7.881665     7.182692     8.499527     7.333424     8.590961     7.253989     9.666757     8.244887     9.152183     7.744161 
  0.30     1.400538     1.356144     1.678072     1.613532     2.298658     2.232661     3.307429     3.307429     4.007196     3.981853     4.769772     4.765535     5.491853     5.361768     5.807355     5.550901     6.372952     6.157044     6.849999     6.565597     7.018812     6.672425     7.516015     6.622930     7.464668     6.531069     8.912650     7.879706     8.056584     7.422233 
  0.35     1.124328     1.097611     1.344828     1.275007     1.516015     1.389567     2.773996     2.731183     3.500802     3.500802     4.389567     4.365972     5.027685     4.930737     5.291309     5.187451     6.047015     5.993674     6.488001     6.339137     5.969012     5.867896     7.232661     6.266037     7.095924     6.277985     7.881665     7.329124     7.089159     6.724650 
  0.40     1.070389     1.042644     1.077243     1.049631     1.263034     1.214125     2.333424     2.321928     3.097611     3.087463     3.819668     3.801159     4.613532     4.611172     5.106013     4.980025     5.384050     5.350497     5.714795     5.529821     5.419539     5.253989     6.863938     6.090853     6.422233     5.531069     7.523562     7.327687     6.198494     6.127633 
  0.45     1.056584     1.056584     0.978196     0.978196     1.090853     1.063503     1.782409     1.757023     2.367371     2.316146     3.204767     3.195348     3.976364     3.963474     4.503349     4.472488     5.017922     5.016140     5.028569     4.921246     4.459432     4.319040     6.028569     5.201634     5.516015     4.613532     6.410070     6.226509     5.472488     5.444932 
  0.50     0.879706     0.879706     0.925999     0.910733     0.985500     0.970854     1.378512     1.367371     2.035624     2.000000     2.292782     2.292782     2.531069     2.521051     3.620586     3.618239     3.885574     3.869871     4.224966     4.209453     3.794936     3.560715     5.722466     4.669027     5.469886     4.650765     6.561937     6.444270     4.075533     4.073820 
  }\plotdata;
		  \begin{axis} [
      axis lines=left,
      xlabel={$\alpha$},
      ylabel={$Log_2$ of samples},
      ymin=0, ymax=15,
      xmin=0.05, xmax=0.5,
      minor x tick num=1,
      minor y tick num=4,
      legend cell align=left,
      legend style={ draw=none, at={(1.35,1.0)}, },
    ]
    \addplot[red]    table[ x index=0, y index=1,  ]                {\plotdata};
    \addplot[cyan]   table[ x index=0, y index=5,  ]                {\plotdata};
    \addplot[violet] table[ x index=0, y index=9,  ]                {\plotdata};
    \addplot[teal]   table[ x index=0, y index=13, ]                {\plotdata};
    \addplot[blue]   table[ x index=0, y index=17, ]                {\plotdata};
    \addplot[orange] table[ x index=0, y index=21, ]                {\plotdata};
    \addplot[yellow] table[ x index=0, y index=25, ]                {\plotdata};
    \addplot[magenta]table[ x index=0, y index=29, ]                {\plotdata};
    \legend{$c = 2$,$c = 4$,$c = 6$,$c = 8$,$c = 10$,$c = 12$,$c = 14$,$c = 16$}
  \end{axis}
	    \end{tikzpicture}
	}
	\caption{ $R=3$ } \label{fig:STSRM03r}
    \end{subfigure}
    \hspace*{\fill} 
    \begin{subfigure}{0.32\textwidth}
	\scalebox{0.48}{
	    \begin{tikzpicture}
		  \pgfplotstableread{
 Alpha       rawL02       mqcL02       rawL03       mqcL03       rawL04       mqcL04       rawL05       mqcL05       rawL06       mqcL06       rawL07       mqcL07       rawL08       mqcL08       rawL09       mqcL09       rawL10       mqcL10       rawL11       mqcL11       rawL12       mqcL12       rawL13       mqcL13       rawL14       mqcL14       rawL15       mqcL15       rawL16       mqcL16
  0.05     0.111031     0.111031     1.541019     1.395063     2.548437     2.386811     2.970854     2.891419     3.720278     3.347666     3.757023     3.513491     5.206331     4.498251     5.107688     4.432959     5.646163     4.712596     6.048759     4.916477     6.681449     5.389567     7.213347     5.215679     8.378512     5.531069     8.620000     5.636915     9.376082     5.942984 
  0.10     1.416840     1.339137     2.748461     2.687061     3.863938     3.650765     4.687061     4.528571     5.134221     4.655352     6.010780     5.280956     6.960697     6.197708     7.641546     6.147307     9.414136     6.839960    10.132577     7.250962    11.089194     7.523562    11.652939     7.518535    12.245817     7.689299    12.633636     7.970854    12.742504     8.403268 
  0.15     1.726831     1.608809     2.871844     2.823749     4.121015     4.028569     5.129283     4.871844     5.693766     5.232661     6.593354     5.693766     8.247928     7.250962     8.523562     6.682573     9.690417     7.289834    10.249445     7.765535    11.002786     8.016140    12.076567     8.493135    12.397847     8.655352    12.605427     9.100978    12.788496     9.272023 
  0.20     1.687061     1.555816     2.344828     2.298658     4.033863     4.000000     4.847997     4.695994     5.454176     5.134221     6.243364     5.563158     7.613532     6.859970     7.978196     6.769772     8.669027     7.272023     9.209453     7.416840     9.799087     7.867896    11.202604     8.104337    11.726105     8.414136    12.271655     9.014355    12.367371     9.100978 
  0.25     1.669027     1.594549     2.550901     2.485427     3.384050     3.319040     4.397803     4.260026     5.070389     4.761285     5.671293     5.253989     6.954196     6.563158     7.232661     6.356144     8.211012     7.107688     8.731183     7.589763     9.316146     7.680324    10.495695     8.214125    11.417592     8.253989    11.551453     8.618239    11.561710     8.653060 
  0.30     1.722466     1.613532     2.163499     2.042644     3.321928     3.275007     4.232661     4.117695     4.780310     4.625270     5.565597     5.214125     6.662205     6.249445     6.853996     6.080658     7.615887     6.922198     8.204767     7.333424     8.560715     7.408712     9.687061     7.541019    10.511595     7.946731     9.692650     8.046142    11.123024     8.680324 
  0.35     1.443607     1.333424     2.084064     1.963474     2.687061     2.599318     3.769772     3.750607     4.480265     4.395063     5.147307     4.925999     6.595742     6.396434     6.752749     6.144046     7.333424     6.865919     7.847997     7.250962     7.742006     7.063503     9.424922     7.650765    10.157044     7.994580     9.283922     8.053111    10.323370     8.448901 
  0.40     1.422233     1.333424     1.985500     1.839960     2.659925     2.555816     3.761285     3.739848     4.584963     4.503349     4.933573     4.794936     6.106013     5.967169     6.370164     5.922198     7.280956     6.735522     7.416840     6.948601     7.367371     6.735522     9.106013     7.526069     9.171527     7.440952     9.046142     7.998196     9.653060     8.250962 
  0.45     1.356144     1.238787     1.847997     1.604071     2.464668     2.356144     3.769772     3.726831     4.250962     4.169925     4.910733     4.729009     6.021480     5.852998     6.364572     5.941106     6.827819     6.469886     7.596935     7.090853     7.110196     6.670161     9.017922     7.485427     9.344828     7.807355     9.119356     8.117695     9.911692     8.759156 
  0.50     1.214125     1.176323     1.695994     1.443607     2.400538     2.244887     3.397803     3.384050     4.053111     3.959770     4.495695     4.414136     6.076388     5.904966     6.275007     5.941106     6.522935     6.377818     7.560715     7.087463     7.264536     6.809414     8.726831     7.370164     9.066950     7.673556     8.867896     7.974529     9.611763     8.559492 
  }\plotdata;
		  \begin{axis} [
      axis lines=left,
      xlabel={$\alpha$},
      ylabel={$Log_2$ of samples},
      ymin=0, ymax=15,
      xmin=0.05, xmax=0.5,
      minor x tick num=1,
      minor y tick num=4,
      legend cell align=left,
      legend style={ draw=none, at={(1.35,1.0)}, },
    ]
    \addplot[red]    table[ x index=0, y index=1,  ]                {\plotdata};
    \addplot[cyan]   table[ x index=0, y index=5,  ]                {\plotdata};
    \addplot[violet] table[ x index=0, y index=9,  ]                {\plotdata};
    \addplot[teal]   table[ x index=0, y index=13, ]                {\plotdata};
    \addplot[blue]   table[ x index=0, y index=17, ]                {\plotdata};
    \addplot[orange] table[ x index=0, y index=21, ]                {\plotdata};
    \addplot[yellow] table[ x index=0, y index=25, ]                {\plotdata};
    \addplot[magenta]table[ x index=0, y index=29, ]                {\plotdata};
    \legend{$c = 2$,$c = 4$,$c = 6$,$c = 8$,$c = 10$,$c = 12$,$c = 14$,$c = 16$}
  \end{axis}
	    \end{tikzpicture}
	}
	\caption{ $R=\infty$ } \label{fig:STSRM99r}
    \end{subfigure}
     \caption{ The number of samples from the D-Wave needed to solve a type 1 frustrated cluster loop problem. }
    \label{fig:STSRr}
\end{figure}
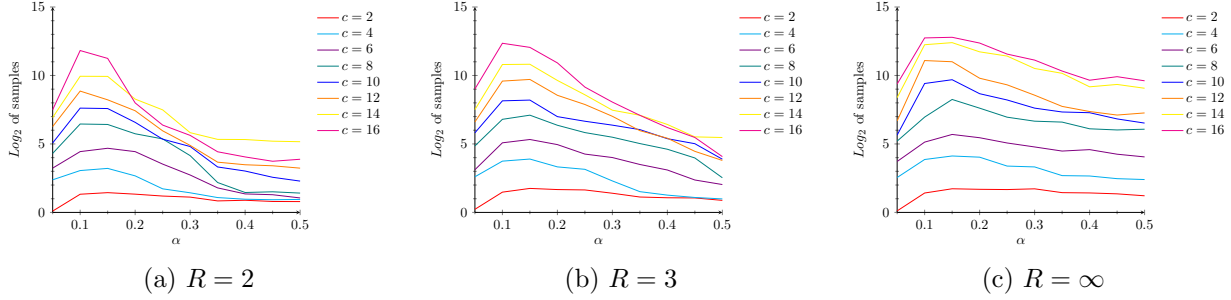

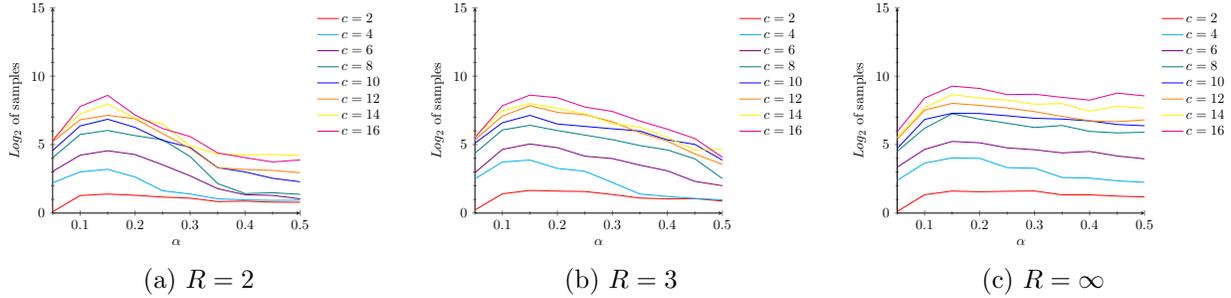
\begin{figure}
    \begin{subfigure}{0.32\textwidth}
	\scalebox{0.48}{
	    \begin{tikzpicture}
		  \pgfplotstableread{
 Alpha       rawL02       mqcL02       rawL03       mqcL03       rawL04       mqcL04       rawL05       mqcL05       rawL06       mqcL06       rawL07       mqcL07       rawL08       mqcL08       rawL09       mqcL09       rawL10       mqcL10       rawL11       mqcL11       rawL12       mqcL12       rawL13       mqcL13       rawL14       mqcL14       rawL15       mqcL15       rawL16       mqcL16
  0.05     0.097611     0.097611     1.691534     1.580145     2.381283     2.217231     2.922198     2.575312     3.235727     3.046142     3.568032     3.364572     4.310340     4.044394     4.627607     4.220330     5.072106     4.578939     5.617063     4.957915     6.272023     5.212569     6.560104     5.407353     6.932628     5.454176     7.349790     5.577731     7.500802     5.275007 
  0.10     1.327687     1.280956     2.709291     2.632268     3.056584     3.007196     4.215679     4.139142     4.440952     4.229588     5.416840     4.944858     6.455492     5.740928     6.771886     5.737687     7.615887     6.364572     8.444932     6.929791     8.859970     6.817623     9.358988     7.165108     9.942045     7.226509    11.057144     7.803227    11.820326     7.788686 
  0.15     1.443607     1.389567     2.622930     2.604071     3.214125     3.195348     4.234195     4.116032     4.684819     4.553361     5.263034     4.980025     6.422233     6.035624     7.118526     6.450221     7.584963     6.851999     7.822730     7.082362     8.221877     7.140779     9.090853     7.401903     9.932886     7.974529    10.375039     8.232661    11.246345     8.596935 
  0.20     1.333424     1.298658     2.084064     2.021480     2.669027     2.632268     3.613532     3.596935     4.438293     4.275007     4.906891     4.713696     5.743084     5.654206     6.283922     5.967169     6.577731     6.257011     6.555816     6.280956     7.424922     6.883621     7.283922     6.454176     8.274262     6.916477     8.399171     7.526069     7.996389     7.157044 
  0.25     1.195348     1.169925     1.613532     1.555816     1.722466     1.632268     2.731183     2.726831     3.528571     3.526069     4.012569     3.942984     5.368070     5.358256     5.681449     5.617063     5.348374     5.302319     4.928844     4.886550     5.957915     5.809414     6.702658     5.773996     7.489286     6.513491     7.412782     7.019702     6.378512     6.174726 
  0.30     1.117695     1.090853     1.275007     1.263034     1.432959     1.389567     1.855990     1.847997     2.731183     2.718088     3.277985     3.266037     4.147307     4.129283     4.733354     4.680324     4.812498     4.795975     3.925999     3.912650     4.893362     4.720278     6.063503     5.201634     5.823749     4.887525     6.742006     6.574102     5.625270     5.589763 
  0.35     0.839960     0.839960     1.007196     0.992768     1.084064     1.042644     1.244887     1.244887     1.790772     1.782409     2.077243     2.063503     2.182692     2.157044     3.650765     3.641546     3.319040     3.319040     3.060047     3.014355     3.669027     3.310340     5.526069     4.632268     5.333424     4.310340     5.653060     5.607626     4.427606     4.393691 
  0.40     0.887525     0.871844     0.855990     0.855990     0.963474     0.963474     1.104337     1.104337     1.350497     1.339137     1.655352     1.655352     1.454176     1.422233     3.080658     3.060047     3.021480     3.003602     2.405992     2.344828     3.480265     3.182692     5.443607     4.495695     5.321928     4.250962     4.811471     4.648465     4.051372     4.047887 
  0.45     0.799087     0.799087     0.863938     0.863938     0.925999     0.925999     1.014355     1.014355     1.292782     1.292782     1.773996     1.765535     1.505891     1.485427     3.049631     3.049631     2.565597     2.541019     2.207893     2.124328     3.405992     3.117695     5.505891     4.604071     5.201634     4.275007     5.063503     4.933573     3.735522     3.733354 
  0.50     0.790772     0.790772     0.823749     0.823749     0.948601     0.948601     1.104337     1.104337     1.056584     1.042644     1.739848     1.731183     1.411426     1.367371     3.395063     3.389567     2.286881     2.286881     2.250962     2.189034     3.232661     2.948601     5.226509     4.137504     5.163499     4.214125     5.304511     5.264536     3.881665     3.879706 
  }\plotdata;
		  \begin{axis} [
      axis lines=left,
      xlabel={$\alpha$},
      ylabel={$Log_2$ of samples},
      ymin=0, ymax=15,
      xmin=0.05, xmax=0.5,
      minor x tick num=1,
      minor y tick num=4,
      legend cell align=left,
      legend style={ draw=none, at={(1.35,1.0)}, },
    ]
    \addplot[red]    table[ x index=0, y index=2,  ]                {\plotdata};
    \addplot[cyan]   table[ x index=0, y index=6,  ]                {\plotdata};
    \addplot[violet] table[ x index=0, y index=10, ]                {\plotdata};
    \addplot[teal]   table[ x index=0, y index=14, ]                {\plotdata};
    \addplot[blue]   table[ x index=0, y index=18, ]                {\plotdata};
    \addplot[orange] table[ x index=0, y index=22, ]                {\plotdata};
    \addplot[yellow] table[ x index=0, y index=26, ]                {\plotdata};
    \addplot[magenta]table[ x index=0, y index=30, ]                {\plotdata};
    \legend{$c = 2$,$c = 4$,$c = 6$,$c = 8$,$c = 10$,$c = 12$,$c = 14$,$c = 16$}
  \end{axis}
	    \end{tikzpicture}
	}
	\caption{ $R=2$ } \label{fig:STSRM02m}
    \end{subfigure}
    \hspace*{\fill} 
    \begin{subfigure}{0.32\textwidth}
	\scalebox{0.48}{
	    \begin{tikzpicture}
		  \pgfplotstableread{
 Alpha       rawL02       mqcL02       rawL03       mqcL03       rawL04       mqcL04       rawL05       mqcL05       rawL06       mqcL06       rawL07       mqcL07       rawL08       mqcL08       rawL09       mqcL09       rawL10       mqcL10       rawL11       mqcL11       rawL12       mqcL12       rawL13       mqcL13       rawL14       mqcL14       rawL15       mqcL15       rawL16       mqcL16
  0.05     0.226509     0.226509     1.613532     1.454176     2.601697     2.503349     3.021480     2.823749     3.117695     2.952334     4.258519     3.748461     4.853996     4.371559     4.746313     4.286881     5.812498     5.102658     6.160275     5.104337     6.607626     5.344828     7.017922     5.153805     7.503349     5.582556     7.747924     5.295723     9.015248     5.516015 
  0.10     1.480265     1.405992     2.669027     2.565597     3.744161     3.729009     4.189034     4.025029     5.082362     4.650765     5.713696     5.152183     6.798051     6.070389     7.142413     5.797013     8.147307     6.599318     9.393004     7.049631     9.589763     7.090853    10.580145     7.456806    10.799975     7.475085    12.208947     7.773996    12.351763     7.839960 
  0.15     1.752749     1.655352     2.678072     2.604071     3.897240     3.873813     4.589763     4.440952     5.324811     5.046142     5.790772     5.432959     7.100978     6.411426     7.363171     6.341986     8.203201     7.137504     9.291309     7.498251     9.710493     7.843984    10.171527     7.851999    10.818760     8.003602    12.208995     8.761285    12.051225     8.615887 
  0.20     1.669027     1.604071     2.469886     2.356144     3.324811     3.263034     4.134221     4.054848     4.952334     4.778209     5.558268     5.316146     6.364572     6.028569     6.925999     6.250962     6.994580     6.505891     7.967169     7.275007     8.552131     7.357552     9.099295     7.587365     9.638074     7.648465    10.941106     8.451541    10.918620     8.422233 
  0.25     1.641546     1.575312     2.130931     2.007196     3.147307     3.056584     3.726831     3.698218     4.260026     4.155425     5.037382     4.942984     5.829850     5.698218     6.087463     5.678072     6.657640     6.330558     7.241840     6.893362     7.881665     7.182692     8.499527     7.333424     8.590961     7.253989     9.666757     8.244887     9.152183     7.744161 
  0.30     1.400538     1.356144     1.678072     1.613532     2.298658     2.232661     3.307429     3.307429     4.007196     3.981853     4.769772     4.765535     5.491853     5.361768     5.807355     5.550901     6.372952     6.157044     6.849999     6.565597     7.018812     6.672425     7.516015     6.622930     7.464668     6.531069     8.912650     7.879706     8.056584     7.422233 
  0.35     1.124328     1.097611     1.344828     1.275007     1.516015     1.389567     2.773996     2.731183     3.500802     3.500802     4.389567     4.365972     5.027685     4.930737     5.291309     5.187451     6.047015     5.993674     6.488001     6.339137     5.969012     5.867896     7.232661     6.266037     7.095924     6.277985     7.881665     7.329124     7.089159     6.724650 
  0.40     1.070389     1.042644     1.077243     1.049631     1.263034     1.214125     2.333424     2.321928     3.097611     3.087463     3.819668     3.801159     4.613532     4.611172     5.106013     4.980025     5.384050     5.350497     5.714795     5.529821     5.419539     5.253989     6.863938     6.090853     6.422233     5.531069     7.523562     7.327687     6.198494     6.127633 
  0.45     1.056584     1.056584     0.978196     0.978196     1.090853     1.063503     1.782409     1.757023     2.367371     2.316146     3.204767     3.195348     3.976364     3.963474     4.503349     4.472488     5.017922     5.016140     5.028569     4.921246     4.459432     4.319040     6.028569     5.201634     5.516015     4.613532     6.410070     6.226509     5.472488     5.444932 
  0.50     0.879706     0.879706     0.925999     0.910733     0.985500     0.970854     1.378512     1.367371     2.035624     2.000000     2.292782     2.292782     2.531069     2.521051     3.620586     3.618239     3.885574     3.869871     4.224966     4.209453     3.794936     3.560715     5.722466     4.669027     5.469886     4.650765     6.561937     6.444270     4.075533     4.073820 
  }\plotdata;
		  \begin{axis} [
      axis lines=left,
      xlabel={$\alpha$},
      ylabel={$Log_2$ of samples},
      ymin=0, ymax=15,
      xmin=0.05, xmax=0.5,
      minor x tick num=1,
      minor y tick num=4,
      legend cell align=left,
      legend style={ draw=none, at={(1.35,1.0)}, },
    ]
    \addplot[red]    table[ x index=0, y index=2,  ]                {\plotdata};
    \addplot[cyan]   table[ x index=0, y index=6,  ]                {\plotdata};
    \addplot[violet] table[ x index=0, y index=10, ]                {\plotdata};
    \addplot[teal]   table[ x index=0, y index=14, ]                {\plotdata};
    \addplot[blue]   table[ x index=0, y index=18, ]                {\plotdata};
    \addplot[orange] table[ x index=0, y index=22, ]                {\plotdata};
    \addplot[yellow] table[ x index=0, y index=26, ]                {\plotdata};
    \addplot[magenta]table[ x index=0, y index=30, ]                {\plotdata};
    \legend{$c = 2$,$c = 4$,$c = 6$,$c = 8$,$c = 10$,$c = 12$,$c = 14$,$c = 16$}
  \end{axis}
	    \end{tikzpicture}
	}
	\caption{ $R=3$ } \label{fig:STSRM03m}
    \end{subfigure}
    \hspace*{\fill} 
    \begin{subfigure}{0.32\textwidth}
	\scalebox{0.48}{
	    \begin{tikzpicture}
		  \pgfplotstableread{
 Alpha       rawL02       mqcL02       rawL03       mqcL03       rawL04       mqcL04       rawL05       mqcL05       rawL06       mqcL06       rawL07       mqcL07       rawL08       mqcL08       rawL09       mqcL09       rawL10       mqcL10       rawL11       mqcL11       rawL12       mqcL12       rawL13       mqcL13       rawL14       mqcL14       rawL15       mqcL15       rawL16       mqcL16
  0.05     0.111031     0.111031     1.541019     1.395063     2.548437     2.386811     2.970854     2.891419     3.720278     3.347666     3.757023     3.513491     5.206331     4.498251     5.107688     4.432959     5.646163     4.712596     6.048759     4.916477     6.681449     5.389567     7.213347     5.215679     8.378512     5.531069     8.620000     5.636915     9.376082     5.942984 
  0.10     1.416840     1.339137     2.748461     2.687061     3.863938     3.650765     4.687061     4.528571     5.134221     4.655352     6.010780     5.280956     6.960697     6.197708     7.641546     6.147307     9.414136     6.839960    10.132577     7.250962    11.089194     7.523562    11.652939     7.518535    12.245817     7.689299    12.633636     7.970854    12.742504     8.403268 
  0.15     1.726831     1.608809     2.871844     2.823749     4.121015     4.028569     5.129283     4.871844     5.693766     5.232661     6.593354     5.693766     8.247928     7.250962     8.523562     6.682573     9.690417     7.289834    10.249445     7.765535    11.002786     8.016140    12.076567     8.493135    12.397847     8.655352    12.605427     9.100978    12.788496     9.272023 
  0.20     1.687061     1.555816     2.344828     2.298658     4.033863     4.000000     4.847997     4.695994     5.454176     5.134221     6.243364     5.563158     7.613532     6.859970     7.978196     6.769772     8.669027     7.272023     9.209453     7.416840     9.799087     7.867896    11.202604     8.104337    11.726105     8.414136    12.271655     9.014355    12.367371     9.100978 
  0.25     1.669027     1.594549     2.550901     2.485427     3.384050     3.319040     4.397803     4.260026     5.070389     4.761285     5.671293     5.253989     6.954196     6.563158     7.232661     6.356144     8.211012     7.107688     8.731183     7.589763     9.316146     7.680324    10.495695     8.214125    11.417592     8.253989    11.551453     8.618239    11.561710     8.653060 
  0.30     1.722466     1.613532     2.163499     2.042644     3.321928     3.275007     4.232661     4.117695     4.780310     4.625270     5.565597     5.214125     6.662205     6.249445     6.853996     6.080658     7.615887     6.922198     8.204767     7.333424     8.560715     7.408712     9.687061     7.541019    10.511595     7.946731     9.692650     8.046142    11.123024     8.680324 
  0.35     1.443607     1.333424     2.084064     1.963474     2.687061     2.599318     3.769772     3.750607     4.480265     4.395063     5.147307     4.925999     6.595742     6.396434     6.752749     6.144046     7.333424     6.865919     7.847997     7.250962     7.742006     7.063503     9.424922     7.650765    10.157044     7.994580     9.283922     8.053111    10.323370     8.448901 
  0.40     1.422233     1.333424     1.985500     1.839960     2.659925     2.555816     3.761285     3.739848     4.584963     4.503349     4.933573     4.794936     6.106013     5.967169     6.370164     5.922198     7.280956     6.735522     7.416840     6.948601     7.367371     6.735522     9.106013     7.526069     9.171527     7.440952     9.046142     7.998196     9.653060     8.250962 
  0.45     1.356144     1.238787     1.847997     1.604071     2.464668     2.356144     3.769772     3.726831     4.250962     4.169925     4.910733     4.729009     6.021480     5.852998     6.364572     5.941106     6.827819     6.469886     7.596935     7.090853     7.110196     6.670161     9.017922     7.485427     9.344828     7.807355     9.119356     8.117695     9.911692     8.759156 
  0.50     1.214125     1.176323     1.695994     1.443607     2.400538     2.244887     3.397803     3.384050     4.053111     3.959770     4.495695     4.414136     6.076388     5.904966     6.275007     5.941106     6.522935     6.377818     7.560715     7.087463     7.264536     6.809414     8.726831     7.370164     9.066950     7.673556     8.867896     7.974529     9.611763     8.559492 
  }\plotdata;
		  \begin{axis} [
      axis lines=left,
      xlabel={$\alpha$},
      ylabel={$Log_2$ of samples},
      ymin=0, ymax=15,
      xmin=0.05, xmax=0.5,
      minor x tick num=1,
      minor y tick num=4,
      legend cell align=left,
      legend style={ draw=none, at={(1.35,1.0)}, },
    ]
    \addplot[red]    table[ x index=0, y index=2,  ]                {\plotdata};
    \addplot[cyan]   table[ x index=0, y index=6,  ]                {\plotdata};
    \addplot[violet] table[ x index=0, y index=10, ]                {\plotdata};
    \addplot[teal]   table[ x index=0, y index=14, ]                {\plotdata};
    \addplot[blue]   table[ x index=0, y index=18, ]                {\plotdata};
    \addplot[orange] table[ x index=0, y index=22, ]                {\plotdata};
    \addplot[yellow] table[ x index=0, y index=26, ]                {\plotdata};
    \addplot[magenta]table[ x index=0, y index=30, ]                {\plotdata};
    \legend{$c = 2$,$c = 4$,$c = 6$,$c = 8$,$c = 10$,$c = 12$,$c = 14$,$c = 16$}
  \end{axis}
	    \end{tikzpicture}
	}
	\caption{ $R=\infty$ } \label{fig:STSRM99m}
    \end{subfigure}
     \caption{ The number of samples from the D-Wave needed for MQC to solve a type 1 frustrated cluster loop problem. }
    \label{fig:STSRMm}
\end{figure}

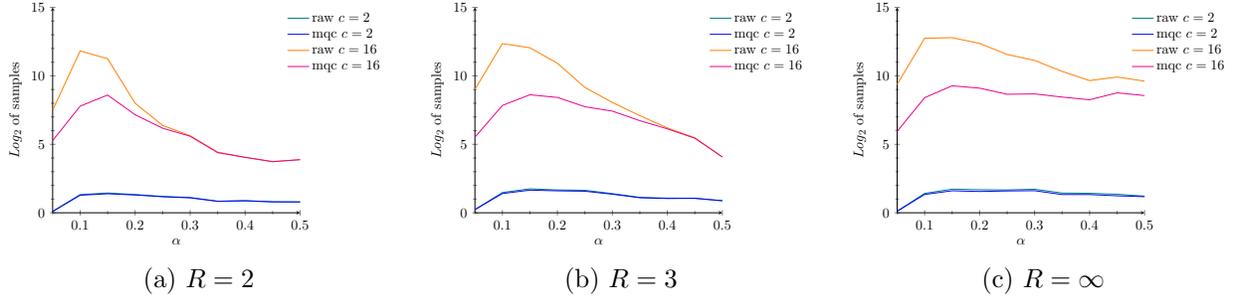
\begin{figure}
    \begin{subfigure}{0.32\textwidth}
	\scalebox{0.48}{
	    \begin{tikzpicture}
		  \pgfplotstableread{
 Alpha       rawL02       mqcL02       rawL03       mqcL03       rawL04       mqcL04       rawL05       mqcL05       rawL06       mqcL06       rawL07       mqcL07       rawL08       mqcL08       rawL09       mqcL09       rawL10       mqcL10       rawL11       mqcL11       rawL12       mqcL12       rawL13       mqcL13       rawL14       mqcL14       rawL15       mqcL15       rawL16       mqcL16
  0.05     0.097611     0.097611     1.691534     1.580145     2.381283     2.217231     2.922198     2.575312     3.235727     3.046142     3.568032     3.364572     4.310340     4.044394     4.627607     4.220330     5.072106     4.578939     5.617063     4.957915     6.272023     5.212569     6.560104     5.407353     6.932628     5.454176     7.349790     5.577731     7.500802     5.275007 
  0.10     1.327687     1.280956     2.709291     2.632268     3.056584     3.007196     4.215679     4.139142     4.440952     4.229588     5.416840     4.944858     6.455492     5.740928     6.771886     5.737687     7.615887     6.364572     8.444932     6.929791     8.859970     6.817623     9.358988     7.165108     9.942045     7.226509    11.057144     7.803227    11.820326     7.788686 
  0.15     1.443607     1.389567     2.622930     2.604071     3.214125     3.195348     4.234195     4.116032     4.684819     4.553361     5.263034     4.980025     6.422233     6.035624     7.118526     6.450221     7.584963     6.851999     7.822730     7.082362     8.221877     7.140779     9.090853     7.401903     9.932886     7.974529    10.375039     8.232661    11.246345     8.596935 
  0.20     1.333424     1.298658     2.084064     2.021480     2.669027     2.632268     3.613532     3.596935     4.438293     4.275007     4.906891     4.713696     5.743084     5.654206     6.283922     5.967169     6.577731     6.257011     6.555816     6.280956     7.424922     6.883621     7.283922     6.454176     8.274262     6.916477     8.399171     7.526069     7.996389     7.157044 
  0.25     1.195348     1.169925     1.613532     1.555816     1.722466     1.632268     2.731183     2.726831     3.528571     3.526069     4.012569     3.942984     5.368070     5.358256     5.681449     5.617063     5.348374     5.302319     4.928844     4.886550     5.957915     5.809414     6.702658     5.773996     7.489286     6.513491     7.412782     7.019702     6.378512     6.174726 
  0.30     1.117695     1.090853     1.275007     1.263034     1.432959     1.389567     1.855990     1.847997     2.731183     2.718088     3.277985     3.266037     4.147307     4.129283     4.733354     4.680324     4.812498     4.795975     3.925999     3.912650     4.893362     4.720278     6.063503     5.201634     5.823749     4.887525     6.742006     6.574102     5.625270     5.589763 
  0.35     0.839960     0.839960     1.007196     0.992768     1.084064     1.042644     1.244887     1.244887     1.790772     1.782409     2.077243     2.063503     2.182692     2.157044     3.650765     3.641546     3.319040     3.319040     3.060047     3.014355     3.669027     3.310340     5.526069     4.632268     5.333424     4.310340     5.653060     5.607626     4.427606     4.393691 
  0.40     0.887525     0.871844     0.855990     0.855990     0.963474     0.963474     1.104337     1.104337     1.350497     1.339137     1.655352     1.655352     1.454176     1.422233     3.080658     3.060047     3.021480     3.003602     2.405992     2.344828     3.480265     3.182692     5.443607     4.495695     5.321928     4.250962     4.811471     4.648465     4.051372     4.047887 
  0.45     0.799087     0.799087     0.863938     0.863938     0.925999     0.925999     1.014355     1.014355     1.292782     1.292782     1.773996     1.765535     1.505891     1.485427     3.049631     3.049631     2.565597     2.541019     2.207893     2.124328     3.405992     3.117695     5.505891     4.604071     5.201634     4.275007     5.063503     4.933573     3.735522     3.733354 
  0.50     0.790772     0.790772     0.823749     0.823749     0.948601     0.948601     1.104337     1.104337     1.056584     1.042644     1.739848     1.731183     1.411426     1.367371     3.395063     3.389567     2.286881     2.286881     2.250962     2.189034     3.232661     2.948601     5.226509     4.137504     5.163499     4.214125     5.304511     5.264536     3.881665     3.879706 
  }\plotdata;
		  \begin{axis} [
      axis lines=left,
      xlabel={$\alpha$},
      ylabel={$Log_2$ of samples},
      ymin=0, ymax=15,
      xmin=0.05, xmax=0.5,
      minor x tick num=1,
      minor y tick num=4,
      legend cell align=left,
      legend style={ draw=none, at={(1.35,1.0)}, },
    ]
    \addplot[teal]   table[ x index=0, y index=1, ]     {\plotdata};
    \addplot[blue]   table[ x index=0, y index=2, ]     {\plotdata};
    \addplot[orange] table[ x index=0, y index=29, ]    {\plotdata};
    \addplot[magenta]table[ x index=0, y index=30, ]    {\plotdata};
    \legend{raw $c = 2$,mqc $c = 2$,raw $c = 16$,mqc $c = 16$}
  \end{axis}
	    \end{tikzpicture}
	}
	\caption{ $R=2$ } \label{fig:STSRM02l}
    \end{subfigure}
    \hspace*{\fill} 
    \begin{subfigure}{0.32\textwidth}
	\scalebox{0.48}{
	    \begin{tikzpicture}
		  \pgfplotstableread{
 Alpha       rawL02       mqcL02       rawL03       mqcL03       rawL04       mqcL04       rawL05       mqcL05       rawL06       mqcL06       rawL07       mqcL07       rawL08       mqcL08       rawL09       mqcL09       rawL10       mqcL10       rawL11       mqcL11       rawL12       mqcL12       rawL13       mqcL13       rawL14       mqcL14       rawL15       mqcL15       rawL16       mqcL16
  0.05     0.226509     0.226509     1.613532     1.454176     2.601697     2.503349     3.021480     2.823749     3.117695     2.952334     4.258519     3.748461     4.853996     4.371559     4.746313     4.286881     5.812498     5.102658     6.160275     5.104337     6.607626     5.344828     7.017922     5.153805     7.503349     5.582556     7.747924     5.295723     9.015248     5.516015 
  0.10     1.480265     1.405992     2.669027     2.565597     3.744161     3.729009     4.189034     4.025029     5.082362     4.650765     5.713696     5.152183     6.798051     6.070389     7.142413     5.797013     8.147307     6.599318     9.393004     7.049631     9.589763     7.090853    10.580145     7.456806    10.799975     7.475085    12.208947     7.773996    12.351763     7.839960 
  0.15     1.752749     1.655352     2.678072     2.604071     3.897240     3.873813     4.589763     4.440952     5.324811     5.046142     5.790772     5.432959     7.100978     6.411426     7.363171     6.341986     8.203201     7.137504     9.291309     7.498251     9.710493     7.843984    10.171527     7.851999    10.818760     8.003602    12.208995     8.761285    12.051225     8.615887 
  0.20     1.669027     1.604071     2.469886     2.356144     3.324811     3.263034     4.134221     4.054848     4.952334     4.778209     5.558268     5.316146     6.364572     6.028569     6.925999     6.250962     6.994580     6.505891     7.967169     7.275007     8.552131     7.357552     9.099295     7.587365     9.638074     7.648465    10.941106     8.451541    10.918620     8.422233 
  0.25     1.641546     1.575312     2.130931     2.007196     3.147307     3.056584     3.726831     3.698218     4.260026     4.155425     5.037382     4.942984     5.829850     5.698218     6.087463     5.678072     6.657640     6.330558     7.241840     6.893362     7.881665     7.182692     8.499527     7.333424     8.590961     7.253989     9.666757     8.244887     9.152183     7.744161 
  0.30     1.400538     1.356144     1.678072     1.613532     2.298658     2.232661     3.307429     3.307429     4.007196     3.981853     4.769772     4.765535     5.491853     5.361768     5.807355     5.550901     6.372952     6.157044     6.849999     6.565597     7.018812     6.672425     7.516015     6.622930     7.464668     6.531069     8.912650     7.879706     8.056584     7.422233 
  0.35     1.124328     1.097611     1.344828     1.275007     1.516015     1.389567     2.773996     2.731183     3.500802     3.500802     4.389567     4.365972     5.027685     4.930737     5.291309     5.187451     6.047015     5.993674     6.488001     6.339137     5.969012     5.867896     7.232661     6.266037     7.095924     6.277985     7.881665     7.329124     7.089159     6.724650 
  0.40     1.070389     1.042644     1.077243     1.049631     1.263034     1.214125     2.333424     2.321928     3.097611     3.087463     3.819668     3.801159     4.613532     4.611172     5.106013     4.980025     5.384050     5.350497     5.714795     5.529821     5.419539     5.253989     6.863938     6.090853     6.422233     5.531069     7.523562     7.327687     6.198494     6.127633 
  0.45     1.056584     1.056584     0.978196     0.978196     1.090853     1.063503     1.782409     1.757023     2.367371     2.316146     3.204767     3.195348     3.976364     3.963474     4.503349     4.472488     5.017922     5.016140     5.028569     4.921246     4.459432     4.319040     6.028569     5.201634     5.516015     4.613532     6.410070     6.226509     5.472488     5.444932 
  0.50     0.879706     0.879706     0.925999     0.910733     0.985500     0.970854     1.378512     1.367371     2.035624     2.000000     2.292782     2.292782     2.531069     2.521051     3.620586     3.618239     3.885574     3.869871     4.224966     4.209453     3.794936     3.560715     5.722466     4.669027     5.469886     4.650765     6.561937     6.444270     4.075533     4.073820 
  }\plotdata;
		  \begin{axis} [
      axis lines=left,
      xlabel={$\alpha$},
      ylabel={$Log_2$ of samples},
      ymin=0, ymax=15,
      xmin=0.05, xmax=0.5,
      minor x tick num=1,
      minor y tick num=4,
      legend cell align=left,
      legend style={ draw=none, at={(1.35,1.0)}, },
    ]
    \addplot[teal]   table[ x index=0, y index=1, ]     {\plotdata};
    \addplot[blue]   table[ x index=0, y index=2, ]     {\plotdata};
    \addplot[orange] table[ x index=0, y index=29, ]    {\plotdata};
    \addplot[magenta]table[ x index=0, y index=30, ]    {\plotdata};
    \legend{raw $c = 2$,mqc $c = 2$,raw $c = 16$,mqc $c = 16$}
  \end{axis}
	    \end{tikzpicture}
	}
	\caption{ $R=3$ } \label{fig:STSRM03l}
    \end{subfigure}
    \hspace*{\fill} 
    \begin{subfigure}{0.32\textwidth}
	\scalebox{0.48}{
	    \begin{tikzpicture}
		  \pgfplotstableread{
 Alpha       rawL02       mqcL02       rawL03       mqcL03       rawL04       mqcL04       rawL05       mqcL05       rawL06       mqcL06       rawL07       mqcL07       rawL08       mqcL08       rawL09       mqcL09       rawL10       mqcL10       rawL11       mqcL11       rawL12       mqcL12       rawL13       mqcL13       rawL14       mqcL14       rawL15       mqcL15       rawL16       mqcL16
  0.05     0.111031     0.111031     1.541019     1.395063     2.548437     2.386811     2.970854     2.891419     3.720278     3.347666     3.757023     3.513491     5.206331     4.498251     5.107688     4.432959     5.646163     4.712596     6.048759     4.916477     6.681449     5.389567     7.213347     5.215679     8.378512     5.531069     8.620000     5.636915     9.376082     5.942984 
  0.10     1.416840     1.339137     2.748461     2.687061     3.863938     3.650765     4.687061     4.528571     5.134221     4.655352     6.010780     5.280956     6.960697     6.197708     7.641546     6.147307     9.414136     6.839960    10.132577     7.250962    11.089194     7.523562    11.652939     7.518535    12.245817     7.689299    12.633636     7.970854    12.742504     8.403268 
  0.15     1.726831     1.608809     2.871844     2.823749     4.121015     4.028569     5.129283     4.871844     5.693766     5.232661     6.593354     5.693766     8.247928     7.250962     8.523562     6.682573     9.690417     7.289834    10.249445     7.765535    11.002786     8.016140    12.076567     8.493135    12.397847     8.655352    12.605427     9.100978    12.788496     9.272023 
  0.20     1.687061     1.555816     2.344828     2.298658     4.033863     4.000000     4.847997     4.695994     5.454176     5.134221     6.243364     5.563158     7.613532     6.859970     7.978196     6.769772     8.669027     7.272023     9.209453     7.416840     9.799087     7.867896    11.202604     8.104337    11.726105     8.414136    12.271655     9.014355    12.367371     9.100978 
  0.25     1.669027     1.594549     2.550901     2.485427     3.384050     3.319040     4.397803     4.260026     5.070389     4.761285     5.671293     5.253989     6.954196     6.563158     7.232661     6.356144     8.211012     7.107688     8.731183     7.589763     9.316146     7.680324    10.495695     8.214125    11.417592     8.253989    11.551453     8.618239    11.561710     8.653060 
  0.30     1.722466     1.613532     2.163499     2.042644     3.321928     3.275007     4.232661     4.117695     4.780310     4.625270     5.565597     5.214125     6.662205     6.249445     6.853996     6.080658     7.615887     6.922198     8.204767     7.333424     8.560715     7.408712     9.687061     7.541019    10.511595     7.946731     9.692650     8.046142    11.123024     8.680324 
  0.35     1.443607     1.333424     2.084064     1.963474     2.687061     2.599318     3.769772     3.750607     4.480265     4.395063     5.147307     4.925999     6.595742     6.396434     6.752749     6.144046     7.333424     6.865919     7.847997     7.250962     7.742006     7.063503     9.424922     7.650765    10.157044     7.994580     9.283922     8.053111    10.323370     8.448901 
  0.40     1.422233     1.333424     1.985500     1.839960     2.659925     2.555816     3.761285     3.739848     4.584963     4.503349     4.933573     4.794936     6.106013     5.967169     6.370164     5.922198     7.280956     6.735522     7.416840     6.948601     7.367371     6.735522     9.106013     7.526069     9.171527     7.440952     9.046142     7.998196     9.653060     8.250962 
  0.45     1.356144     1.238787     1.847997     1.604071     2.464668     2.356144     3.769772     3.726831     4.250962     4.169925     4.910733     4.729009     6.021480     5.852998     6.364572     5.941106     6.827819     6.469886     7.596935     7.090853     7.110196     6.670161     9.017922     7.485427     9.344828     7.807355     9.119356     8.117695     9.911692     8.759156 
  0.50     1.214125     1.176323     1.695994     1.443607     2.400538     2.244887     3.397803     3.384050     4.053111     3.959770     4.495695     4.414136     6.076388     5.904966     6.275007     5.941106     6.522935     6.377818     7.560715     7.087463     7.264536     6.809414     8.726831     7.370164     9.066950     7.673556     8.867896     7.974529     9.611763     8.559492 
  }\plotdata;
		  \begin{axis} [
      axis lines=left,
      xlabel={$\alpha$},
      ylabel={$Log_2$ of samples},
      ymin=0, ymax=15,
      xmin=0.05, xmax=0.5,
      minor x tick num=1,
      minor y tick num=4,
      legend cell align=left,
      legend style={ draw=none, at={(1.35,1.0)}, },
    ]
    \addplot[teal]   table[ x index=0, y index=1, ]     {\plotdata};
    \addplot[blue]   table[ x index=0, y index=2, ]     {\plotdata};
    \addplot[orange] table[ x index=0, y index=29, ]    {\plotdata};
    \addplot[magenta]table[ x index=0, y index=30, ]    {\plotdata};
    \legend{raw $c = 2$,mqc $c = 2$,raw $c = 16$,mqc $c = 16$}
  \end{axis}
	    \end{tikzpicture}
	}
	\caption{ $R=\infty$ } \label{fig:STSRM99l}
    \end{subfigure}
     \caption{ The number of samples from the D-Wave needed to solve a type 1 frustrated cluster loop problem. 
     Comparing only D-Wave sizes of $c=2$ ($L_c = 32$) and $c=16$ ($L_c = 2048$) . }
    \label{fig:STSRMl}
\end{figure}

Figure \ref{fig:STSRr} (D-Wave alone) and figure \ref{fig:STSRMm} (D-Wave w/ MQC) present the results for 
type 1 FCL problems as a family of curves where each line represents 
a different size partial D-Wave from a 2x2 cell partition to the full 2048 qubit, 16x16 cell D-Wave.
When the problem is easier, such as a 2x2 cell partition or when $\alpha$ is between 0.3 and 0.5, 
MQC post-processing gives very little improvement over the D-Wave results alone.
However when the problems get harder, when using the entire D-Wave and $\alpha$ is around 0.1, 
MQC post-processing gives as much as a 10 to 20 times improvement over the D-Wave alone.
Figure \ref{fig:STSRMl} was added to show a clearer distinction between the behavior of the D-Wave alone 
and the D-Wave with MQC post-processing on type 1 FCL on a full 2048 qubit D-Wave. 
Note that there is no difference in the behavior for a 32 ($c=2$) qubit partial D-Wave.

\begin{figure}
    \begin{subfigure}{0.32\textwidth}
	\scalebox{0.48}{
	    \begin{tikzpicture}
		  \pgfplotstableread{
 Alpha       rawL02       mqcL02       rawL03       mqcL03       rawL04       mqcL04       rawL05       mqcL05       rawL06       mqcL06       rawL07       mqcL07       rawL08       mqcL08       rawL09       mqcL09       rawL10       mqcL10       rawL11       mqcL11       rawL12       mqcL12       rawL13       mqcL13       rawL14       mqcL14       rawL15       mqcL15       rawL16       mqcL16 
  0.05     0.070389     0.070389     1.367371     1.214125     2.107688     1.981853     2.500802     2.310340     2.422233     2.207893     2.992768     2.871844     3.495695     3.195348     4.085765     3.797013     4.324811     3.985500     4.422233     3.990955     4.589763     3.989139     5.129283     4.288359     5.127633     4.336283     5.488001     4.438293     6.319040     4.809414 
  0.10     1.310340     1.250962     2.117695     2.021480     3.253989     3.114367     3.813525     3.709291     4.232661     4.039138     4.722466     4.381283     5.367371     5.111031     6.107688     5.580145     6.712596     5.538538     7.363872     5.865919     7.541019     5.729009     8.619757     6.211012     9.376969     6.518535    10.118111     6.485427    10.771523     6.851999 
  0.15     1.545968     1.475085     2.459432     2.395063     3.389567     3.327687     4.179511     4.073820     4.746313     4.485427     5.592158     5.277985     6.408712     6.054848     7.434295     6.462052     8.000901     6.950468     8.779260     7.117695     9.576522     7.214125    10.049631     7.615887    11.100394     8.089159    11.790991     8.446256    11.441372     8.235727 
  0.20     1.485427     1.411426     2.130931     2.021480     3.584963     3.518535     4.039138     3.942984     4.599318     4.471187     5.187451     4.972693     6.343408     6.200065     7.107688     6.518535     7.769772     7.053111     7.978196     6.952334     8.782409     7.422233     9.998196     7.803227    10.769772     8.217231    10.156096     8.150560     9.838249     8.025029 
  0.25     1.176323     1.150560     1.799087     1.722466     2.757023     2.673556     3.448901     3.395063     4.078951     4.040892     4.788686     4.613532     5.923149     5.882643     6.388190     6.261531     7.100978     6.823749     7.488001     6.996389     7.709291     7.028569     8.922198     7.422233     9.778209     7.718088     8.584963     7.618239     7.514753     6.845992 
  0.30     1.372952     1.361768     1.459432     1.427606     2.070389     2.014355     2.673556     2.650765     3.389567     3.375735     4.182692     4.129283     5.249445     5.211012     5.881665     5.742006     6.414136     6.260026     6.516015     6.333424     7.147307     6.634593     8.169925     7.049631     8.769772     7.536053     7.471187     6.987321     6.742006     6.542258 
  0.35     1.021480     1.007196     1.007196     1.007196     1.310340     1.238787     2.201634     2.189034     2.914565     2.899176     3.488001     3.482848     4.524816     4.516015     5.711495     5.634593     5.897240     5.825786     6.287620     6.144862     6.499527     6.320485     7.375735     6.451541     8.403268     7.303050     6.221877     5.987321     5.622930     5.456806 
  0.40     0.985500     0.970854     1.000000     1.000000     1.137504     1.097611     1.384050     1.384050     1.827819     1.827819     2.438293     2.432959     3.443607     3.432959     4.390943     4.390943     4.933573     4.929791     5.478972     5.357552     5.472488     5.218781     6.500802     5.641546     7.341986     6.378512     5.147307     5.046142     4.553361     4.526069 
  0.45     0.871844     0.871844     0.956057     0.941106     1.014355     1.000000     1.189034     1.163499     1.405992     1.405992     1.847997     1.839960     1.594549     1.565597     3.495695     3.482848     4.166715     4.153805     4.742006     4.729009     4.780310     4.709291     5.823749     4.903038     6.257011     5.327687     4.508429     4.438293     3.356144     3.339137 
  0.50     0.895303     0.895303     0.985500     0.970854     0.963474     0.963474     1.084064     1.084064     1.344828     1.344828     1.713696     1.704872     1.604071     1.555816     2.903038     2.887525     4.026800     4.026800     4.612352     4.596935     4.477677     4.367371     5.823749     4.925999     6.304511     5.531069     4.253989     4.104337     3.330558     3.307429 
  }\plotdata;
		  \begin{axis} [
      axis lines=left,
      xlabel={$\alpha$},
      ylabel={$Log_2$ of samples},
      ymin=0, ymax=15,
      xmin=0.05, xmax=0.5,
      minor x tick num=1,
      minor y tick num=4,
      legend cell align=left,
      legend style={ draw=none, at={(1.35,1.0)}, },
    ]
    \addplot[red]    table[ x index=0, y index=1,  ]                {\plotdata};
    \addplot[cyan]   table[ x index=0, y index=5,  ]                {\plotdata};
    \addplot[violet] table[ x index=0, y index=9,  ]                {\plotdata};
    \addplot[teal]   table[ x index=0, y index=13, ]                {\plotdata};
    \addplot[blue]   table[ x index=0, y index=17, ]                {\plotdata};
    \addplot[orange] table[ x index=0, y index=21, ]                {\plotdata};
    \addplot[yellow] table[ x index=0, y index=25, ]                {\plotdata};
    \addplot[magenta]table[ x index=0, y index=29, ]                {\plotdata};
    \legend{$c = 2$,$c = 4$,$c = 6$,$c = 8$,$c = 10$,$c = 12$,$c = 14$,$c = 16$}
  \end{axis}
	    \end{tikzpicture}
	}
	\caption{ $R=2$ } \label{fig:STSRC02r}
    \end{subfigure}
    \hspace*{\fill} 
    \begin{subfigure}{0.32\textwidth}
	\scalebox{0.48}{
	    \begin{tikzpicture}
		  \pgfplotstableread{
 Alpha       rawL02       mqcL02       rawL03       mqcL03       rawL04       mqcL04       rawL05       mqcL05       rawL06       mqcL06       rawL07       mqcL07       rawL08       mqcL08       rawL09       mqcL09       rawL10       mqcL10       rawL11       mqcL11       rawL12       mqcL12       rawL13       mqcL13       rawL14       mqcL14       rawL15       mqcL15       rawL16       mqcL16 
  0.05     0.111031     0.111031     1.286881     1.163499     2.266037     2.039138     2.400538     2.286881     2.765535     2.636915     2.948601     2.769772     3.871844     3.475085     4.132577     3.669027     4.275007     3.731183     4.980025     4.272023     4.920293     4.223423     5.572890     4.316146     5.695994     4.500802     6.152995     4.784504     6.533563     4.695994 
  0.10     1.384050     1.316146     2.475085     2.405992     3.378512     3.169925     4.039138     3.831877     4.543496     4.300124     5.135863     4.648465     5.969012     5.364572     7.101818     6.000000     7.311794     5.776104     8.084064     6.134221     8.514122     5.935460     9.852947     6.283922    10.573991     6.490570    11.421259     6.948601    11.365476     7.273516 
  0.15     1.604071     1.485427     2.604071     2.526069     3.700440     3.646163     4.462052     4.321928     5.152183     4.867896     5.880686     5.288359     6.837943     6.316146     7.937344     6.908813     8.404631     6.625270     9.538538     7.356144    10.493135     7.416840    11.517013     7.757023    12.085465     8.240314    12.274262     8.232661    12.404596     8.541019 
  0.20     1.815575     1.765535     2.416840     2.367371     3.414136     3.392317     4.389567     4.292782     5.063503     4.843984     5.908813     5.500802     6.998196     6.558268     8.032101     7.260026     8.198494     6.735522     9.424922     7.843984     9.757023     7.613532    10.970556     7.992768    11.701645     8.321928    11.651732     8.646163    11.173127     8.319040 
  0.25     1.526069     1.485427     2.327687     2.220330     3.204767     3.153805     4.040892     3.980025     4.724650     4.563158     5.361768     5.130931     6.350497     6.070389     7.310340     6.985500     7.289834     6.526069     8.408712     7.459432     9.010780     7.505891    10.300124     7.761285    10.837943     8.124328    10.030639     7.992768     9.130931     7.700440 
  0.30     1.604071     1.555816     2.007196     1.941106     2.589763     2.510962     3.799087     3.769772     4.283922     4.253989     5.028569     4.891419     6.193772     6.061776     7.136684     6.878725     6.837943     6.482848     7.788686     7.226509     8.253989     7.195348     9.386811     7.622930     9.956057     7.910733     8.718088     7.687061     7.935460     7.198494 
  0.35     1.333424     1.310340     1.443607     1.321928     2.432959     2.316146     3.438293     3.411426     3.967169     3.906891     4.577731     4.536053     5.854993     5.764474     6.803227     6.691534     6.553361     6.253989     7.316146     6.970854     7.847997     7.049631     8.941106     7.333424     9.467279     7.675816     8.014355     7.341986     7.627607     7.117695 
  0.40     1.084064     1.042644     1.201634     1.163499     1.678072     1.584963     3.066950     3.035624     3.643856     3.643856     4.147307     4.127633     5.121015     5.080658     6.212569     6.161888     6.229588     5.969012     7.037382     6.809414     7.392317     6.843984     8.829850     7.367371     8.983678     7.439623     7.220330     6.765535     7.488001     7.058316 
  0.45     0.992768     0.992768     1.263034     1.189034     1.526069     1.356144     2.372952     2.344828     3.220330     3.217231     3.998196     3.996389     4.786596     4.767655     5.974529     5.924100     5.920293     5.841973     6.715893     6.584963     7.195348     6.937344     8.116032     6.931683     8.201634     7.028569     6.505891     6.247928     7.375735     7.114367 
  0.50     0.933573     0.933573     1.063503     1.035624     1.130931     1.077243     1.978196     1.963474     2.803227     2.782409     3.541019     3.533563     4.353323     4.326250     5.221877     5.139142     5.462052     5.367371     6.132577     6.066950     6.253989     6.033863     7.996389     6.929791     7.881665     6.790772     5.784504     5.646163     6.754353     6.692092 
  }\plotdata;
		  \begin{axis} [
      axis lines=left,
      xlabel={$\alpha$},
      ylabel={$Log_2$ of samples},
      ymin=0, ymax=15,
      xmin=0.05, xmax=0.5,
      minor x tick num=1,
      minor y tick num=4,
      legend cell align=left,
      legend style={ draw=none, at={(1.35,1.0)}, },
    ]
    \addplot[red]    table[ x index=0, y index=1,  ]                {\plotdata};
    \addplot[cyan]   table[ x index=0, y index=5,  ]                {\plotdata};
    \addplot[violet] table[ x index=0, y index=9,  ]                {\plotdata};
    \addplot[teal]   table[ x index=0, y index=13, ]                {\plotdata};
    \addplot[blue]   table[ x index=0, y index=17, ]                {\plotdata};
    \addplot[orange] table[ x index=0, y index=21, ]                {\plotdata};
    \addplot[yellow] table[ x index=0, y index=25, ]                {\plotdata};
    \addplot[magenta]table[ x index=0, y index=29, ]                {\plotdata};
    \legend{$c = 2$,$c = 4$,$c = 6$,$c = 8$,$c = 10$,$c = 12$,$c = 14$,$c = 16$}
  \end{axis}
	    \end{tikzpicture}
	}
	\caption{ $R=3$ } \label{fig:STSRC03r}
    \end{subfigure}
    \hspace*{\fill} 
    \begin{subfigure}{0.32\textwidth}
	\scalebox{0.48}{
	    \begin{tikzpicture}
		  \pgfplotstableread{
 Alpha       rawL02       mqcL02       rawL03       mqcL03       rawL04       mqcL04       rawL05       mqcL05       rawL06       mqcL06       rawL07       mqcL07       rawL08       mqcL08       rawL09       mqcL09       rawL10       mqcL10       rawL11       mqcL11       rawL12       mqcL12       rawL13       mqcL13       rawL14       mqcL14       rawL15       mqcL15       rawL16       mqcL16 
  0.05     0.111031     0.111031     1.250962     1.124328     2.157044     1.970854     2.372952     2.195348     3.028569     2.859970     3.173127     2.875780     3.929791     3.655352     4.565597     3.883621     4.746313     3.922198     5.097611     4.354734     5.395063     4.370164     5.224966     4.286881     6.514122     4.704872     6.897240     5.046142     7.132577     4.914565 
  0.10     1.389567     1.286881     2.298658     2.182692     3.558268     3.427606     3.873813     3.682573     4.797013     4.343408     5.684819     4.903038     6.452859     5.598127     7.269033     5.851999     7.827819     5.761285     8.972693     6.339137     9.625082     6.153805    10.348281     6.277985    10.783291     6.456806    11.408582     7.000000    12.021402     6.985500 
  0.15     1.485427     1.400538     2.709291     2.659925     3.904966     3.782409     4.693766     4.378512     5.634593     5.157044     6.282440     5.528571     7.694880     6.662205     8.849999     6.827819     9.488286     6.843984    10.841645     7.698218    11.491171     7.604071    12.043301     7.849999    12.475733     8.367371    12.780219     8.786596    12.862496     8.859970 
  0.20     1.739848     1.704872     2.682573     2.627607     3.851999     3.729009     4.857981     4.657640     5.538538     5.253989     6.662205     5.963474     8.047887     6.996389     8.857981     7.207893     9.862947     7.189034    10.666412     7.925999    11.563901     8.169925    12.174029     8.232661    12.467126     8.687061    12.736966     9.195348    12.687324     8.922198 
  0.25     1.695994     1.650765     2.613532     2.500802     3.704872     3.659925     4.475085     4.372952     5.395063     5.169925     6.298658     5.731183     7.321928     6.867896     8.372952     7.063503     8.895303     7.021480     9.904966     7.778209    10.468675     7.863938    11.772194     8.100978    12.321928     8.395063    12.411814     8.875780    11.760501     8.769772 
  0.30     1.669027     1.622930     2.384050     2.275007     3.673556     3.553361     4.277985     4.142413     5.229588     4.952334     5.744161     5.454176     7.258519     6.751678     7.521051     6.910733     8.053111     6.827819     9.087463     7.350497    10.010780     7.956057    11.169925     7.903038    11.805113     8.195348    11.981853     8.739848    10.224966     8.381283 
  0.35     1.713696     1.650765     2.339137     2.137504     3.211012     3.153805     4.114367     4.025029     4.695994     4.580145     5.689299     5.277985     6.996389     6.636915     7.327687     6.608809     7.655352     6.709291     8.464668     7.111031     9.459432     7.700440    10.440952     7.584963    10.959358     7.981853    10.606658     8.269033    11.607802     9.153805 
  0.40     1.286881     1.226509     2.021480     1.863938     2.925999     2.879706     3.877744     3.845992     4.778209     4.664483     5.361768     5.124328     6.386811     6.155425     6.974529     6.459432     7.275007     6.485427     8.053111     6.967169     8.875780     7.735522     9.817623     7.659925    10.678072     7.895303     9.870858     8.060047    11.647315     9.140779 
  0.45     1.333424     1.275007     1.992768     1.790772     2.704872     2.570463     3.666757     3.592158     4.424922     4.381283     5.207893     4.996389     6.503349     6.171527     6.807355     6.341986     7.039138     6.269033     7.673556     6.790772     8.650765     7.536053     9.395063     7.182692    10.462052     7.823749     9.773996     8.144046    10.665264     8.695994 
  0.50     1.310340     1.238787     1.831877     1.613532     2.459432     2.310340     3.659925     3.582556     4.310340     4.266037     4.959770     4.731183     6.427606     6.168321     6.443607     6.214125     6.918386     6.344828     7.298658     6.570463     8.395063     7.500802     8.996389     6.985500    10.056584     7.704872     9.286881     7.895303    10.448901     8.580145 
  }\plotdata;
		  \begin{axis} [
      axis lines=left,
      xlabel={$\alpha$},
      ylabel={$Log_2$ of samples},
      ymin=0, ymax=15,
      xmin=0.05, xmax=0.5,
      minor x tick num=1,
      minor y tick num=4,
      legend cell align=left,
      legend style={ draw=none, at={(1.35,1.0)}, },
    ]
    \addplot[red]    table[ x index=0, y index=1,  ]                {\plotdata};
    \addplot[cyan]   table[ x index=0, y index=5,  ]                {\plotdata};
    \addplot[violet] table[ x index=0, y index=9,  ]                {\plotdata};
    \addplot[teal]   table[ x index=0, y index=13, ]                {\plotdata};
    \addplot[blue]   table[ x index=0, y index=17, ]                {\plotdata};
    \addplot[orange] table[ x index=0, y index=21, ]                {\plotdata};
    \addplot[yellow] table[ x index=0, y index=25, ]                {\plotdata};
    \addplot[magenta]table[ x index=0, y index=29, ]                {\plotdata};
    \legend{$c = 2$,$c = 4$,$c = 6$,$c = 8$,$c = 10$,$c = 12$,$c = 14$,$c = 16$}
  \end{axis}
	    \end{tikzpicture}
	}
	\caption{ $R=\infty$ } \label{fig:STSRC992r}
    \end{subfigure}
     \caption{ The number of samples from the D-Wave needed to solve a type 2 frustrated cluster loop problem. }
    \label{fig:STSRCr}
\end{figure}
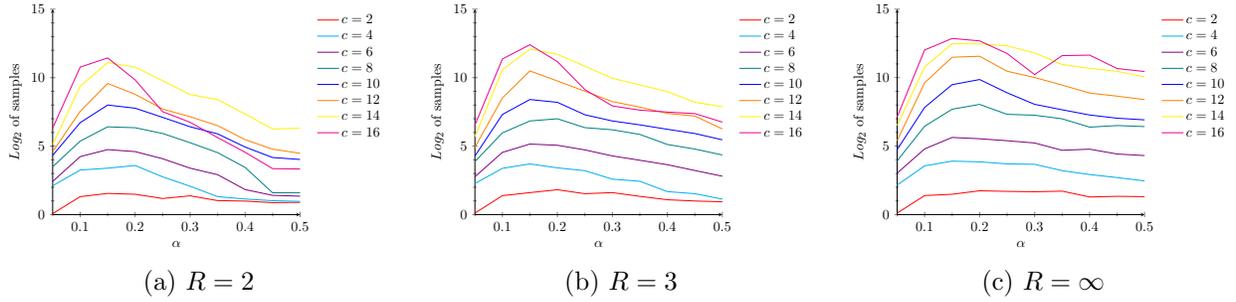

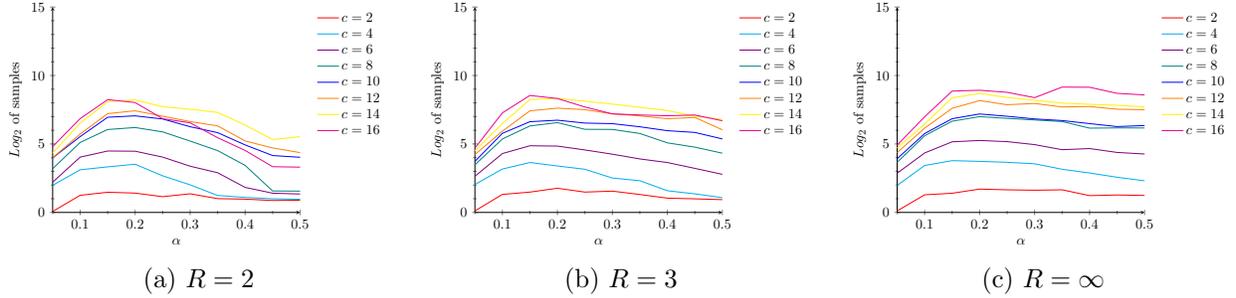
\begin{figure}
    \begin{subfigure}{0.32\textwidth}
	\scalebox{0.48}{
	    \begin{tikzpicture}
		  \pgfplotstableread{
 Alpha       rawL02       mqcL02       rawL03       mqcL03       rawL04       mqcL04       rawL05       mqcL05       rawL06       mqcL06       rawL07       mqcL07       rawL08       mqcL08       rawL09       mqcL09       rawL10       mqcL10       rawL11       mqcL11       rawL12       mqcL12       rawL13       mqcL13       rawL14       mqcL14       rawL15       mqcL15       rawL16       mqcL16 
  0.05     0.070389     0.070389     1.367371     1.214125     2.107688     1.981853     2.500802     2.310340     2.422233     2.207893     2.992768     2.871844     3.495695     3.195348     4.085765     3.797013     4.324811     3.985500     4.422233     3.990955     4.589763     3.989139     5.129283     4.288359     5.127633     4.336283     5.488001     4.438293     6.319040     4.809414 
  0.10     1.310340     1.250962     2.117695     2.021480     3.253989     3.114367     3.813525     3.709291     4.232661     4.039138     4.722466     4.381283     5.367371     5.111031     6.107688     5.580145     6.712596     5.538538     7.363872     5.865919     7.541019     5.729009     8.619757     6.211012     9.376969     6.518535    10.118111     6.485427    10.771523     6.851999 
  0.15     1.545968     1.475085     2.459432     2.395063     3.389567     3.327687     4.179511     4.073820     4.746313     4.485427     5.592158     5.277985     6.408712     6.054848     7.434295     6.462052     8.000901     6.950468     8.779260     7.117695     9.576522     7.214125    10.049631     7.615887    11.100394     8.089159    11.790991     8.446256    11.441372     8.235727 
  0.20     1.485427     1.411426     2.130931     2.021480     3.584963     3.518535     4.039138     3.942984     4.599318     4.471187     5.187451     4.972693     6.343408     6.200065     7.107688     6.518535     7.769772     7.053111     7.978196     6.952334     8.782409     7.422233     9.998196     7.803227    10.769772     8.217231    10.156096     8.150560     9.838249     8.025029 
  0.25     1.176323     1.150560     1.799087     1.722466     2.757023     2.673556     3.448901     3.395063     4.078951     4.040892     4.788686     4.613532     5.923149     5.882643     6.388190     6.261531     7.100978     6.823749     7.488001     6.996389     7.709291     7.028569     8.922198     7.422233     9.778209     7.718088     8.584963     7.618239     7.514753     6.845992 
  0.30     1.372952     1.361768     1.459432     1.427606     2.070389     2.014355     2.673556     2.650765     3.389567     3.375735     4.182692     4.129283     5.249445     5.211012     5.881665     5.742006     6.414136     6.260026     6.516015     6.333424     7.147307     6.634593     8.169925     7.049631     8.769772     7.536053     7.471187     6.987321     6.742006     6.542258 
  0.35     1.021480     1.007196     1.007196     1.007196     1.310340     1.238787     2.201634     2.189034     2.914565     2.899176     3.488001     3.482848     4.524816     4.516015     5.711495     5.634593     5.897240     5.825786     6.287620     6.144862     6.499527     6.320485     7.375735     6.451541     8.403268     7.303050     6.221877     5.987321     5.622930     5.456806 
  0.40     0.985500     0.970854     1.000000     1.000000     1.137504     1.097611     1.384050     1.384050     1.827819     1.827819     2.438293     2.432959     3.443607     3.432959     4.390943     4.390943     4.933573     4.929791     5.478972     5.357552     5.472488     5.218781     6.500802     5.641546     7.341986     6.378512     5.147307     5.046142     4.553361     4.526069 
  0.45     0.871844     0.871844     0.956057     0.941106     1.014355     1.000000     1.189034     1.163499     1.405992     1.405992     1.847997     1.839960     1.594549     1.565597     3.495695     3.482848     4.166715     4.153805     4.742006     4.729009     4.780310     4.709291     5.823749     4.903038     6.257011     5.327687     4.508429     4.438293     3.356144     3.339137 
  0.50     0.895303     0.895303     0.985500     0.970854     0.963474     0.963474     1.084064     1.084064     1.344828     1.344828     1.713696     1.704872     1.604071     1.555816     2.903038     2.887525     4.026800     4.026800     4.612352     4.596935     4.477677     4.367371     5.823749     4.925999     6.304511     5.531069     4.253989     4.104337     3.330558     3.307429 
  }\plotdata;
		  \begin{axis} [
      axis lines=left,
      xlabel={$\alpha$},
      ylabel={$Log_2$ of samples},
      ymin=0, ymax=15,
      xmin=0.05, xmax=0.5,
      minor x tick num=1,
      minor y tick num=4,
      legend cell align=left,
      legend style={ draw=none, at={(1.35,1.0)}, },
    ]
    \addplot[red]    table[ x index=0, y index=2,  ]                {\plotdata};
    \addplot[cyan]   table[ x index=0, y index=6,  ]                {\plotdata};
    \addplot[violet] table[ x index=0, y index=10, ]                {\plotdata};
    \addplot[teal]   table[ x index=0, y index=14, ]                {\plotdata};
    \addplot[blue]   table[ x index=0, y index=18, ]                {\plotdata};
    \addplot[orange] table[ x index=0, y index=22, ]                {\plotdata};
    \addplot[yellow] table[ x index=0, y index=26, ]                {\plotdata};
    \addplot[magenta]table[ x index=0, y index=30, ]                {\plotdata};
    \legend{$c = 2$,$c = 4$,$c = 6$,$c = 8$,$c = 10$,$c = 12$,$c = 14$,$c = 16$}
  \end{axis}
	    \end{tikzpicture}
	}
	\caption{ $R=2$ } \label{fig:STSRC02m}
    \end{subfigure}
    \hspace*{\fill} 
    \begin{subfigure}{0.32\textwidth}
	\scalebox{0.48}{
	    \begin{tikzpicture}
		  \pgfplotstableread{
 Alpha       rawL02       mqcL02       rawL03       mqcL03       rawL04       mqcL04       rawL05       mqcL05       rawL06       mqcL06       rawL07       mqcL07       rawL08       mqcL08       rawL09       mqcL09       rawL10       mqcL10       rawL11       mqcL11       rawL12       mqcL12       rawL13       mqcL13       rawL14       mqcL14       rawL15       mqcL15       rawL16       mqcL16 
  0.05     0.111031     0.111031     1.286881     1.163499     2.266037     2.039138     2.400538     2.286881     2.765535     2.636915     2.948601     2.769772     3.871844     3.475085     4.132577     3.669027     4.275007     3.731183     4.980025     4.272023     4.920293     4.223423     5.572890     4.316146     5.695994     4.500802     6.152995     4.784504     6.533563     4.695994 
  0.10     1.384050     1.316146     2.475085     2.405992     3.378512     3.169925     4.039138     3.831877     4.543496     4.300124     5.135863     4.648465     5.969012     5.364572     7.101818     6.000000     7.311794     5.776104     8.084064     6.134221     8.514122     5.935460     9.852947     6.283922    10.573991     6.490570    11.421259     6.948601    11.365476     7.273516 
  0.15     1.604071     1.485427     2.604071     2.526069     3.700440     3.646163     4.462052     4.321928     5.152183     4.867896     5.880686     5.288359     6.837943     6.316146     7.937344     6.908813     8.404631     6.625270     9.538538     7.356144    10.493135     7.416840    11.517013     7.757023    12.085465     8.240314    12.274262     8.232661    12.404596     8.541019 
  0.20     1.815575     1.765535     2.416840     2.367371     3.414136     3.392317     4.389567     4.292782     5.063503     4.843984     5.908813     5.500802     6.998196     6.558268     8.032101     7.260026     8.198494     6.735522     9.424922     7.843984     9.757023     7.613532    10.970556     7.992768    11.701645     8.321928    11.651732     8.646163    11.173127     8.319040 
  0.25     1.526069     1.485427     2.327687     2.220330     3.204767     3.153805     4.040892     3.980025     4.724650     4.563158     5.361768     5.130931     6.350497     6.070389     7.310340     6.985500     7.289834     6.526069     8.408712     7.459432     9.010780     7.505891    10.300124     7.761285    10.837943     8.124328    10.030639     7.992768     9.130931     7.700440 
  0.30     1.604071     1.555816     2.007196     1.941106     2.589763     2.510962     3.799087     3.769772     4.283922     4.253989     5.028569     4.891419     6.193772     6.061776     7.136684     6.878725     6.837943     6.482848     7.788686     7.226509     8.253989     7.195348     9.386811     7.622930     9.956057     7.910733     8.718088     7.687061     7.935460     7.198494 
  0.35     1.333424     1.310340     1.443607     1.321928     2.432959     2.316146     3.438293     3.411426     3.967169     3.906891     4.577731     4.536053     5.854993     5.764474     6.803227     6.691534     6.553361     6.253989     7.316146     6.970854     7.847997     7.049631     8.941106     7.333424     9.467279     7.675816     8.014355     7.341986     7.627607     7.117695 
  0.40     1.084064     1.042644     1.201634     1.163499     1.678072     1.584963     3.066950     3.035624     3.643856     3.643856     4.147307     4.127633     5.121015     5.080658     6.212569     6.161888     6.229588     5.969012     7.037382     6.809414     7.392317     6.843984     8.829850     7.367371     8.983678     7.439623     7.220330     6.765535     7.488001     7.058316 
  0.45     0.992768     0.992768     1.263034     1.189034     1.526069     1.356144     2.372952     2.344828     3.220330     3.217231     3.998196     3.996389     4.786596     4.767655     5.974529     5.924100     5.920293     5.841973     6.715893     6.584963     7.195348     6.937344     8.116032     6.931683     8.201634     7.028569     6.505891     6.247928     7.375735     7.114367 
  0.50     0.933573     0.933573     1.063503     1.035624     1.130931     1.077243     1.978196     1.963474     2.803227     2.782409     3.541019     3.533563     4.353323     4.326250     5.221877     5.139142     5.462052     5.367371     6.132577     6.066950     6.253989     6.033863     7.996389     6.929791     7.881665     6.790772     5.784504     5.646163     6.754353     6.692092 
  }\plotdata;
		  \begin{axis} [
      axis lines=left,
      xlabel={$\alpha$},
      ylabel={$Log_2$ of samples},
      ymin=0, ymax=15,
      xmin=0.05, xmax=0.5,
      minor x tick num=1,
      minor y tick num=4,
      legend cell align=left,
      legend style={ draw=none, at={(1.35,1.0)}, },
    ]
    \addplot[red]    table[ x index=0, y index=2,  ]                {\plotdata};
    \addplot[cyan]   table[ x index=0, y index=6,  ]                {\plotdata};
    \addplot[violet] table[ x index=0, y index=10, ]                {\plotdata};
    \addplot[teal]   table[ x index=0, y index=14, ]                {\plotdata};
    \addplot[blue]   table[ x index=0, y index=18, ]                {\plotdata};
    \addplot[orange] table[ x index=0, y index=22, ]                {\plotdata};
    \addplot[yellow] table[ x index=0, y index=26, ]                {\plotdata};
    \addplot[magenta]table[ x index=0, y index=30, ]                {\plotdata};
    \legend{$c = 2$,$c = 4$,$c = 6$,$c = 8$,$c = 10$,$c = 12$,$c = 14$,$c = 16$}
  \end{axis}
	    \end{tikzpicture}
	}
	\caption{ $R=3$ } \label{fig:STSRC03m}
    \end{subfigure}
    \hspace*{\fill} 
    \begin{subfigure}{0.32\textwidth}
	\scalebox{0.48}{
	    \begin{tikzpicture}
		  \pgfplotstableread{
 Alpha       rawL02       mqcL02       rawL03       mqcL03       rawL04       mqcL04       rawL05       mqcL05       rawL06       mqcL06       rawL07       mqcL07       rawL08       mqcL08       rawL09       mqcL09       rawL10       mqcL10       rawL11       mqcL11       rawL12       mqcL12       rawL13       mqcL13       rawL14       mqcL14       rawL15       mqcL15       rawL16       mqcL16 
  0.05     0.111031     0.111031     1.250962     1.124328     2.157044     1.970854     2.372952     2.195348     3.028569     2.859970     3.173127     2.875780     3.929791     3.655352     4.565597     3.883621     4.746313     3.922198     5.097611     4.354734     5.395063     4.370164     5.224966     4.286881     6.514122     4.704872     6.897240     5.046142     7.132577     4.914565 
  0.10     1.389567     1.286881     2.298658     2.182692     3.558268     3.427606     3.873813     3.682573     4.797013     4.343408     5.684819     4.903038     6.452859     5.598127     7.269033     5.851999     7.827819     5.761285     8.972693     6.339137     9.625082     6.153805    10.348281     6.277985    10.783291     6.456806    11.408582     7.000000    12.021402     6.985500 
  0.15     1.485427     1.400538     2.709291     2.659925     3.904966     3.782409     4.693766     4.378512     5.634593     5.157044     6.282440     5.528571     7.694880     6.662205     8.849999     6.827819     9.488286     6.843984    10.841645     7.698218    11.491171     7.604071    12.043301     7.849999    12.475733     8.367371    12.780219     8.786596    12.862496     8.859970 
  0.20     1.739848     1.704872     2.682573     2.627607     3.851999     3.729009     4.857981     4.657640     5.538538     5.253989     6.662205     5.963474     8.047887     6.996389     8.857981     7.207893     9.862947     7.189034    10.666412     7.925999    11.563901     8.169925    12.174029     8.232661    12.467126     8.687061    12.736966     9.195348    12.687324     8.922198 
  0.25     1.695994     1.650765     2.613532     2.500802     3.704872     3.659925     4.475085     4.372952     5.395063     5.169925     6.298658     5.731183     7.321928     6.867896     8.372952     7.063503     8.895303     7.021480     9.904966     7.778209    10.468675     7.863938    11.772194     8.100978    12.321928     8.395063    12.411814     8.875780    11.760501     8.769772 
  0.30     1.669027     1.622930     2.384050     2.275007     3.673556     3.553361     4.277985     4.142413     5.229588     4.952334     5.744161     5.454176     7.258519     6.751678     7.521051     6.910733     8.053111     6.827819     9.087463     7.350497    10.010780     7.956057    11.169925     7.903038    11.805113     8.195348    11.981853     8.739848    10.224966     8.381283 
  0.35     1.713696     1.650765     2.339137     2.137504     3.211012     3.153805     4.114367     4.025029     4.695994     4.580145     5.689299     5.277985     6.996389     6.636915     7.327687     6.608809     7.655352     6.709291     8.464668     7.111031     9.459432     7.700440    10.440952     7.584963    10.959358     7.981853    10.606658     8.269033    11.607802     9.153805 
  0.40     1.286881     1.226509     2.021480     1.863938     2.925999     2.879706     3.877744     3.845992     4.778209     4.664483     5.361768     5.124328     6.386811     6.155425     6.974529     6.459432     7.275007     6.485427     8.053111     6.967169     8.875780     7.735522     9.817623     7.659925    10.678072     7.895303     9.870858     8.060047    11.647315     9.140779 
  0.45     1.333424     1.275007     1.992768     1.790772     2.704872     2.570463     3.666757     3.592158     4.424922     4.381283     5.207893     4.996389     6.503349     6.171527     6.807355     6.341986     7.039138     6.269033     7.673556     6.790772     8.650765     7.536053     9.395063     7.182692    10.462052     7.823749     9.773996     8.144046    10.665264     8.695994 
  0.50     1.310340     1.238787     1.831877     1.613532     2.459432     2.310340     3.659925     3.582556     4.310340     4.266037     4.959770     4.731183     6.427606     6.168321     6.443607     6.214125     6.918386     6.344828     7.298658     6.570463     8.395063     7.500802     8.996389     6.985500    10.056584     7.704872     9.286881     7.895303    10.448901     8.580145 
  }\plotdata;
		  \begin{axis} [
      axis lines=left,
      xlabel={$\alpha$},
      ylabel={$Log_2$ of samples},
      ymin=0, ymax=15,
      xmin=0.05, xmax=0.5,
      minor x tick num=1,
      minor y tick num=4,
      legend cell align=left,
      legend style={ draw=none, at={(1.35,1.0)}, },
    ]
    \addplot[red]    table[ x index=0, y index=2,  ]                {\plotdata};
    \addplot[cyan]   table[ x index=0, y index=6,  ]                {\plotdata};
    \addplot[violet] table[ x index=0, y index=10, ]                {\plotdata};
    \addplot[teal]   table[ x index=0, y index=14, ]                {\plotdata};
    \addplot[blue]   table[ x index=0, y index=18, ]                {\plotdata};
    \addplot[orange] table[ x index=0, y index=22, ]                {\plotdata};
    \addplot[yellow] table[ x index=0, y index=26, ]                {\plotdata};
    \addplot[magenta]table[ x index=0, y index=30, ]                {\plotdata};
    \legend{$c = 2$,$c = 4$,$c = 6$,$c = 8$,$c = 10$,$c = 12$,$c = 14$,$c = 16$}
  \end{axis}
	    \end{tikzpicture}
	}
	\caption{ $R=\infty$ } \label{fig:STSRC99m}
    \end{subfigure}
     \caption{ The number of samples from the D-Wave needed for MQC to solve a type 2 frustrated cluster loop problem. }
    \label{fig:STSRCm}
\end{figure}

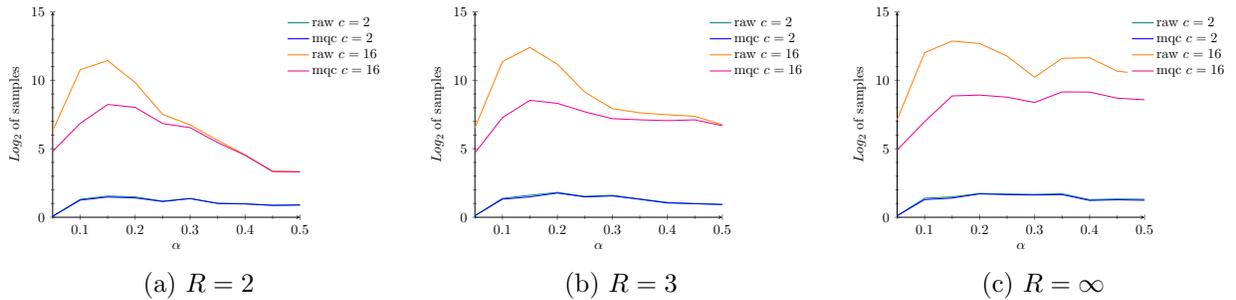
\begin{figure}
    \begin{subfigure}{0.32\textwidth}
	\scalebox{0.48}{
	    \begin{tikzpicture}
		  \pgfplotstableread{
 Alpha       rawL02       mqcL02       rawL03       mqcL03       rawL04       mqcL04       rawL05       mqcL05       rawL06       mqcL06       rawL07       mqcL07       rawL08       mqcL08       rawL09       mqcL09       rawL10       mqcL10       rawL11       mqcL11       rawL12       mqcL12       rawL13       mqcL13       rawL14       mqcL14       rawL15       mqcL15       rawL16       mqcL16 
  0.05     0.070389     0.070389     1.367371     1.214125     2.107688     1.981853     2.500802     2.310340     2.422233     2.207893     2.992768     2.871844     3.495695     3.195348     4.085765     3.797013     4.324811     3.985500     4.422233     3.990955     4.589763     3.989139     5.129283     4.288359     5.127633     4.336283     5.488001     4.438293     6.319040     4.809414 
  0.10     1.310340     1.250962     2.117695     2.021480     3.253989     3.114367     3.813525     3.709291     4.232661     4.039138     4.722466     4.381283     5.367371     5.111031     6.107688     5.580145     6.712596     5.538538     7.363872     5.865919     7.541019     5.729009     8.619757     6.211012     9.376969     6.518535    10.118111     6.485427    10.771523     6.851999 
  0.15     1.545968     1.475085     2.459432     2.395063     3.389567     3.327687     4.179511     4.073820     4.746313     4.485427     5.592158     5.277985     6.408712     6.054848     7.434295     6.462052     8.000901     6.950468     8.779260     7.117695     9.576522     7.214125    10.049631     7.615887    11.100394     8.089159    11.790991     8.446256    11.441372     8.235727 
  0.20     1.485427     1.411426     2.130931     2.021480     3.584963     3.518535     4.039138     3.942984     4.599318     4.471187     5.187451     4.972693     6.343408     6.200065     7.107688     6.518535     7.769772     7.053111     7.978196     6.952334     8.782409     7.422233     9.998196     7.803227    10.769772     8.217231    10.156096     8.150560     9.838249     8.025029 
  0.25     1.176323     1.150560     1.799087     1.722466     2.757023     2.673556     3.448901     3.395063     4.078951     4.040892     4.788686     4.613532     5.923149     5.882643     6.388190     6.261531     7.100978     6.823749     7.488001     6.996389     7.709291     7.028569     8.922198     7.422233     9.778209     7.718088     8.584963     7.618239     7.514753     6.845992 
  0.30     1.372952     1.361768     1.459432     1.427606     2.070389     2.014355     2.673556     2.650765     3.389567     3.375735     4.182692     4.129283     5.249445     5.211012     5.881665     5.742006     6.414136     6.260026     6.516015     6.333424     7.147307     6.634593     8.169925     7.049631     8.769772     7.536053     7.471187     6.987321     6.742006     6.542258 
  0.35     1.021480     1.007196     1.007196     1.007196     1.310340     1.238787     2.201634     2.189034     2.914565     2.899176     3.488001     3.482848     4.524816     4.516015     5.711495     5.634593     5.897240     5.825786     6.287620     6.144862     6.499527     6.320485     7.375735     6.451541     8.403268     7.303050     6.221877     5.987321     5.622930     5.456806 
  0.40     0.985500     0.970854     1.000000     1.000000     1.137504     1.097611     1.384050     1.384050     1.827819     1.827819     2.438293     2.432959     3.443607     3.432959     4.390943     4.390943     4.933573     4.929791     5.478972     5.357552     5.472488     5.218781     6.500802     5.641546     7.341986     6.378512     5.147307     5.046142     4.553361     4.526069 
  0.45     0.871844     0.871844     0.956057     0.941106     1.014355     1.000000     1.189034     1.163499     1.405992     1.405992     1.847997     1.839960     1.594549     1.565597     3.495695     3.482848     4.166715     4.153805     4.742006     4.729009     4.780310     4.709291     5.823749     4.903038     6.257011     5.327687     4.508429     4.438293     3.356144     3.339137 
  0.50     0.895303     0.895303     0.985500     0.970854     0.963474     0.963474     1.084064     1.084064     1.344828     1.344828     1.713696     1.704872     1.604071     1.555816     2.903038     2.887525     4.026800     4.026800     4.612352     4.596935     4.477677     4.367371     5.823749     4.925999     6.304511     5.531069     4.253989     4.104337     3.330558     3.307429 
  }\plotdata;
		  \begin{axis} [
      axis lines=left,
      xlabel={$\alpha$},
      ylabel={$Log_2$ of samples},
      ymin=0, ymax=15,
      xmin=0.05, xmax=0.5,
      minor x tick num=1,
      minor y tick num=4,
      legend cell align=left,
      legend style={ draw=none, at={(1.35,1.0)}, },
    ]
    \addplot[teal]   table[ x index=0, y index=1, ]     {\plotdata};
    \addplot[blue]   table[ x index=0, y index=2, ]     {\plotdata};
    \addplot[orange] table[ x index=0, y index=29, ]    {\plotdata};
    \addplot[magenta]table[ x index=0, y index=30, ]    {\plotdata};
    \legend{raw $c = 2$,mqc $c = 2$,raw $c = 16$,mqc $c = 16$}
  \end{axis}
	    \end{tikzpicture}
	}
	\caption{ $R=2$ } \label{fig:STSRC02l}
    \end{subfigure}
    \hspace*{\fill} 
    \begin{subfigure}{0.32\textwidth}
	\scalebox{0.48}{
	    \begin{tikzpicture}
		  \pgfplotstableread{
 Alpha       rawL02       mqcL02       rawL03       mqcL03       rawL04       mqcL04       rawL05       mqcL05       rawL06       mqcL06       rawL07       mqcL07       rawL08       mqcL08       rawL09       mqcL09       rawL10       mqcL10       rawL11       mqcL11       rawL12       mqcL12       rawL13       mqcL13       rawL14       mqcL14       rawL15       mqcL15       rawL16       mqcL16 
  0.05     0.111031     0.111031     1.286881     1.163499     2.266037     2.039138     2.400538     2.286881     2.765535     2.636915     2.948601     2.769772     3.871844     3.475085     4.132577     3.669027     4.275007     3.731183     4.980025     4.272023     4.920293     4.223423     5.572890     4.316146     5.695994     4.500802     6.152995     4.784504     6.533563     4.695994 
  0.10     1.384050     1.316146     2.475085     2.405992     3.378512     3.169925     4.039138     3.831877     4.543496     4.300124     5.135863     4.648465     5.969012     5.364572     7.101818     6.000000     7.311794     5.776104     8.084064     6.134221     8.514122     5.935460     9.852947     6.283922    10.573991     6.490570    11.421259     6.948601    11.365476     7.273516 
  0.15     1.604071     1.485427     2.604071     2.526069     3.700440     3.646163     4.462052     4.321928     5.152183     4.867896     5.880686     5.288359     6.837943     6.316146     7.937344     6.908813     8.404631     6.625270     9.538538     7.356144    10.493135     7.416840    11.517013     7.757023    12.085465     8.240314    12.274262     8.232661    12.404596     8.541019 
  0.20     1.815575     1.765535     2.416840     2.367371     3.414136     3.392317     4.389567     4.292782     5.063503     4.843984     5.908813     5.500802     6.998196     6.558268     8.032101     7.260026     8.198494     6.735522     9.424922     7.843984     9.757023     7.613532    10.970556     7.992768    11.701645     8.321928    11.651732     8.646163    11.173127     8.319040 
  0.25     1.526069     1.485427     2.327687     2.220330     3.204767     3.153805     4.040892     3.980025     4.724650     4.563158     5.361768     5.130931     6.350497     6.070389     7.310340     6.985500     7.289834     6.526069     8.408712     7.459432     9.010780     7.505891    10.300124     7.761285    10.837943     8.124328    10.030639     7.992768     9.130931     7.700440 
  0.30     1.604071     1.555816     2.007196     1.941106     2.589763     2.510962     3.799087     3.769772     4.283922     4.253989     5.028569     4.891419     6.193772     6.061776     7.136684     6.878725     6.837943     6.482848     7.788686     7.226509     8.253989     7.195348     9.386811     7.622930     9.956057     7.910733     8.718088     7.687061     7.935460     7.198494 
  0.35     1.333424     1.310340     1.443607     1.321928     2.432959     2.316146     3.438293     3.411426     3.967169     3.906891     4.577731     4.536053     5.854993     5.764474     6.803227     6.691534     6.553361     6.253989     7.316146     6.970854     7.847997     7.049631     8.941106     7.333424     9.467279     7.675816     8.014355     7.341986     7.627607     7.117695 
  0.40     1.084064     1.042644     1.201634     1.163499     1.678072     1.584963     3.066950     3.035624     3.643856     3.643856     4.147307     4.127633     5.121015     5.080658     6.212569     6.161888     6.229588     5.969012     7.037382     6.809414     7.392317     6.843984     8.829850     7.367371     8.983678     7.439623     7.220330     6.765535     7.488001     7.058316 
  0.45     0.992768     0.992768     1.263034     1.189034     1.526069     1.356144     2.372952     2.344828     3.220330     3.217231     3.998196     3.996389     4.786596     4.767655     5.974529     5.924100     5.920293     5.841973     6.715893     6.584963     7.195348     6.937344     8.116032     6.931683     8.201634     7.028569     6.505891     6.247928     7.375735     7.114367 
  0.50     0.933573     0.933573     1.063503     1.035624     1.130931     1.077243     1.978196     1.963474     2.803227     2.782409     3.541019     3.533563     4.353323     4.326250     5.221877     5.139142     5.462052     5.367371     6.132577     6.066950     6.253989     6.033863     7.996389     6.929791     7.881665     6.790772     5.784504     5.646163     6.754353     6.692092 
  }\plotdata;
		  \begin{axis} [
      axis lines=left,
      xlabel={$\alpha$},
      ylabel={$Log_2$ of samples},
      ymin=0, ymax=15,
      xmin=0.05, xmax=0.5,
      minor x tick num=1,
      minor y tick num=4,
      legend cell align=left,
      legend style={ draw=none, at={(1.35,1.0)}, },
    ]
    \addplot[teal]   table[ x index=0, y index=1, ]     {\plotdata};
    \addplot[blue]   table[ x index=0, y index=2, ]     {\plotdata};
    \addplot[orange] table[ x index=0, y index=29, ]    {\plotdata};
    \addplot[magenta]table[ x index=0, y index=30, ]    {\plotdata};
    \legend{raw $c = 2$,mqc $c = 2$,raw $c = 16$,mqc $c = 16$}
  \end{axis}
	    \end{tikzpicture}
	}
	\caption{ $R=3$ } \label{fig:STSRC03l}
    \end{subfigure}
    \hspace*{\fill} 
    \begin{subfigure}{0.32\textwidth}
	\scalebox{0.48}{
	    \begin{tikzpicture}
		  \pgfplotstableread{
 Alpha       rawL02       mqcL02       rawL03       mqcL03       rawL04       mqcL04       rawL05       mqcL05       rawL06       mqcL06       rawL07       mqcL07       rawL08       mqcL08       rawL09       mqcL09       rawL10       mqcL10       rawL11       mqcL11       rawL12       mqcL12       rawL13       mqcL13       rawL14       mqcL14       rawL15       mqcL15       rawL16       mqcL16 
  0.05     0.111031     0.111031     1.250962     1.124328     2.157044     1.970854     2.372952     2.195348     3.028569     2.859970     3.173127     2.875780     3.929791     3.655352     4.565597     3.883621     4.746313     3.922198     5.097611     4.354734     5.395063     4.370164     5.224966     4.286881     6.514122     4.704872     6.897240     5.046142     7.132577     4.914565 
  0.10     1.389567     1.286881     2.298658     2.182692     3.558268     3.427606     3.873813     3.682573     4.797013     4.343408     5.684819     4.903038     6.452859     5.598127     7.269033     5.851999     7.827819     5.761285     8.972693     6.339137     9.625082     6.153805    10.348281     6.277985    10.783291     6.456806    11.408582     7.000000    12.021402     6.985500 
  0.15     1.485427     1.400538     2.709291     2.659925     3.904966     3.782409     4.693766     4.378512     5.634593     5.157044     6.282440     5.528571     7.694880     6.662205     8.849999     6.827819     9.488286     6.843984    10.841645     7.698218    11.491171     7.604071    12.043301     7.849999    12.475733     8.367371    12.780219     8.786596    12.862496     8.859970 
  0.20     1.739848     1.704872     2.682573     2.627607     3.851999     3.729009     4.857981     4.657640     5.538538     5.253989     6.662205     5.963474     8.047887     6.996389     8.857981     7.207893     9.862947     7.189034    10.666412     7.925999    11.563901     8.169925    12.174029     8.232661    12.467126     8.687061    12.736966     9.195348    12.687324     8.922198 
  0.25     1.695994     1.650765     2.613532     2.500802     3.704872     3.659925     4.475085     4.372952     5.395063     5.169925     6.298658     5.731183     7.321928     6.867896     8.372952     7.063503     8.895303     7.021480     9.904966     7.778209    10.468675     7.863938    11.772194     8.100978    12.321928     8.395063    12.411814     8.875780    11.760501     8.769772 
  0.30     1.669027     1.622930     2.384050     2.275007     3.673556     3.553361     4.277985     4.142413     5.229588     4.952334     5.744161     5.454176     7.258519     6.751678     7.521051     6.910733     8.053111     6.827819     9.087463     7.350497    10.010780     7.956057    11.169925     7.903038    11.805113     8.195348    11.981853     8.739848    10.224966     8.381283 
  0.35     1.713696     1.650765     2.339137     2.137504     3.211012     3.153805     4.114367     4.025029     4.695994     4.580145     5.689299     5.277985     6.996389     6.636915     7.327687     6.608809     7.655352     6.709291     8.464668     7.111031     9.459432     7.700440    10.440952     7.584963    10.959358     7.981853    10.606658     8.269033    11.607802     9.153805 
  0.40     1.286881     1.226509     2.021480     1.863938     2.925999     2.879706     3.877744     3.845992     4.778209     4.664483     5.361768     5.124328     6.386811     6.155425     6.974529     6.459432     7.275007     6.485427     8.053111     6.967169     8.875780     7.735522     9.817623     7.659925    10.678072     7.895303     9.870858     8.060047    11.647315     9.140779 
  0.45     1.333424     1.275007     1.992768     1.790772     2.704872     2.570463     3.666757     3.592158     4.424922     4.381283     5.207893     4.996389     6.503349     6.171527     6.807355     6.341986     7.039138     6.269033     7.673556     6.790772     8.650765     7.536053     9.395063     7.182692    10.462052     7.823749     9.773996     8.144046    10.665264     8.695994 
  0.50     1.310340     1.238787     1.831877     1.613532     2.459432     2.310340     3.659925     3.582556     4.310340     4.266037     4.959770     4.731183     6.427606     6.168321     6.443607     6.214125     6.918386     6.344828     7.298658     6.570463     8.395063     7.500802     8.996389     6.985500    10.056584     7.704872     9.286881     7.895303    10.448901     8.580145 
  }\plotdata;
		  \begin{axis} [
      axis lines=left,
      xlabel={$\alpha$},
      ylabel={$Log_2$ of samples},
      ymin=0, ymax=15,
      xmin=0.05, xmax=0.5,
      minor x tick num=1,
      minor y tick num=4,
      legend cell align=left,
      legend style={ draw=none, at={(1.35,1.0)}, },
    ]
    \addplot[teal]   table[ x index=0, y index=1, ]     {\plotdata};
    \addplot[blue]   table[ x index=0, y index=2, ]     {\plotdata};
    \addplot[orange] table[ x index=0, y index=29, ]    {\plotdata};
    \addplot[magenta]table[ x index=0, y index=30, ]    {\plotdata};
    \legend{raw $c = 2$,mqc $c = 2$,raw $c = 16$,mqc $c = 16$}
  \end{axis}
	    \end{tikzpicture}
	}
	\caption{ $R=\infty$ } \label{fig:STSRC99l}
    \end{subfigure}
     \caption{ The number of samples from the D-Wave needed to solve a type 2 frustrated cluster loop problem. 
     Comparing only D-Wave sizes of $c=2$ ($L_c = 32$) and $c=16$ ($L_c = 2048$) . }
    \label{fig:STSRCl}
\end{figure}

Figure \ref{fig:STSRCr} (D-Wave alone) and figure \ref{fig:STSRCm} (D-Wave w/ MQC) present the results for 
type 2 FCL problems.
The results are similar to those of the type 1 results.
That is, if the problems are easy, MQC shows very little improvement but if the problem is hard MQC may find the
answer 10 to 20 times faster.
Figure \ref{fig:STSRCl} was added for the same reason Figure \ref{fig:STSRCr} was, to show more clearly the 
differences in the behavior of a full 2048 qubit D-Wave on type 2 FCL problems.

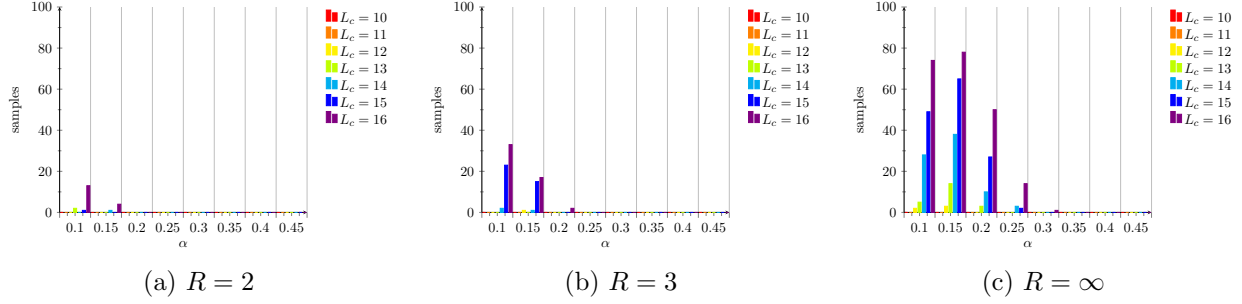
\begin{figure}
    \begin{subfigure}{0.32\textwidth}
	\scalebox{0.48}{
	    \begin{tikzpicture}
		  \pgfplotstableread{
 Alpha       rawL02       rawL03       rawL04       rawL05       rawL06       rawL07       rawL08       rawL09       rawL10       rawL11       rawL12       rawL13       rawL14       rawL15       rawL16 
  0.05            0            0            0            0            0            0            0            0            0            0            0            0            0            0            0 
  0.10            0            0            0            0            0            0            0            0            0            0            0            2            0            1           13 
  0.15            0            0            0            0            0            0            0            0            0            0            0            0            1            0            4 
  0.20            0            0            0            0            0            0            0            0            0            0            0            0            0            0            0 
  0.25            0            0            0            0            0            0            0            0            0            0            0            0            0            0            0 
  0.30            0            0            0            0            0            0            0            0            0            0            0            0            0            0            0 
  0.35            0            0            0            0            0            0            0            0            0            0            0            0            0            0            0 
  0.40            0            0            0            0            0            0            0            0            0            0            0            0            0            0            0 
  0.45            0            0            0            0            0            0            0            0            0            0            0            0            0            0            0 
  0.50            0            0            0            0            0            0            0            0            0            0            0            0            0            0            0 
  }\plotdata;
		  \begin{axis} [
      axis lines=left,
      xlabel={$\alpha$},
      ylabel={samples},
      ymin=0, ymax=100,
      xmin=0.1, xmax=0.5,
      minor x tick num=3,
      minor y tick num=1,
      legend cell align=left,
      legend style={ draw=none, at={(1.35,1.0)}, },
      ybar interval=0.7,
    ]
    \addplot[red,fill=red]       table[ x index=0, y index=9,  ]                {\plotdata};
    \addplot[orange,fill=orange] table[ x index=0, y index=10, ]                {\plotdata};
    \addplot[yellow,fill=yellow] table[ x index=0, y index=11, ]                {\plotdata};
    \addplot[lime,fill=lime]     table[ x index=0, y index=12, ]                {\plotdata};
    \addplot[cyan,fill=cyan]     table[ x index=0, y index=13, ]                {\plotdata};
    \addplot[blue,fill=blue]     table[ x index=0, y index=14, ]                {\plotdata};
    \addplot[violet,fill=violet] table[ x index=0, y index=15, ]                {\plotdata};
    \legend{$L_c = 10$,$L_c = 11$,$L_c = 12$,$L_c = 13$,$L_c = 14$,$L_c = 15$,$L_c = 16$}
  \end{axis}
	    \end{tikzpicture}
	}
	\caption{ $R=2$ } \label{fig:NSTSM2}
    \end{subfigure}
    \hspace*{\fill} 
    \begin{subfigure}{0.32\textwidth}
	\scalebox{0.48}{
	    \begin{tikzpicture}
		  \pgfplotstableread{
 Alpha       rawL02       rawL03       rawL04       rawL05       rawL06       rawL07       rawL08       rawL09       rawL10       rawL11       rawL12       rawL13       rawL14       rawL15       rawL16 
  0.05            0            0            0            0            0            0            0            0            0            0            0            0            0            0            0 
  0.10            0            0            0            0            0            0            0            0            0            0            0            0            2           23           33 
  0.15            0            0            0            0            0            0            0            0            0            0            1            0            1           15           17 
  0.20            0            0            0            0            0            0            0            0            0            0            0            0            0            0            2 
  0.25            0            0            0            0            0            0            0            0            0            0            0            0            0            0            0 
  0.30            0            0            0            0            0            0            0            0            0            0            0            0            0            0            0 
  0.35            0            0            0            0            0            0            0            0            0            0            0            0            0            0            0 
  0.40            0            0            0            0            0            0            0            0            0            0            0            0            0            0            0 
  0.45            0            0            0            0            0            0            0            0            0            0            0            0            0            0            0 
  0.50            0            0            0            0            0            0            0            0            0            0            0            0            0            0            0 
  }\plotdata;
		  \begin{axis} [
      axis lines=left,
      xlabel={$\alpha$},
      ylabel={samples},
      ymin=0, ymax=100,
      xmin=0.1, xmax=0.5,
      minor x tick num=3,
      minor y tick num=1,
      legend cell align=left,
      legend style={ draw=none, at={(1.35,1.0)}, },
      ybar interval=0.7,
    ]
    \addplot[red,fill=red]       table[ x index=0, y index=9,  ]                {\plotdata};
    \addplot[orange,fill=orange] table[ x index=0, y index=10, ]                {\plotdata};
    \addplot[yellow,fill=yellow] table[ x index=0, y index=11, ]                {\plotdata};
    \addplot[lime,fill=lime]     table[ x index=0, y index=12, ]                {\plotdata};
    \addplot[cyan,fill=cyan]     table[ x index=0, y index=13, ]                {\plotdata};
    \addplot[blue,fill=blue]     table[ x index=0, y index=14, ]                {\plotdata};
    \addplot[violet,fill=violet] table[ x index=0, y index=15, ]                {\plotdata};
    \legend{$L_c = 10$,$L_c = 11$,$L_c = 12$,$L_c = 13$,$L_c = 14$,$L_c = 15$,$L_c = 16$}
  \end{axis}
	    \end{tikzpicture}
	}
	\caption{ $R=3$ } \label{fig:NSTSM3}
    \end{subfigure}
    \hspace*{\fill} 
    \begin{subfigure}{0.32\textwidth}
	\scalebox{0.48}{
	    \begin{tikzpicture}
		  \pgfplotstableread{
 Alpha       rawL02       rawL03       rawL04       rawL05       rawL06       rawL07       rawL08       rawL09       rawL10       rawL11       rawL12       rawL13       rawL14       rawL15       rawL16
  0.05            0            0            0            0            0            0            0            0            0            0            0            0            0            0            0 
  0.10            0            0            0            0            0            0            0            0            0            0            2            5           28           49           74 
  0.15            0            0            0            0            0            0            0            0            0            0            3           14           38           65           78 
  0.20            0            0            0            0            0            0            0            0            0            0            0            3           10           27           50 
  0.25            0            0            0            0            0            0            0            0            0            0            0            0            3            2           14 
  0.30            0            0            0            0            0            0            0            0            0            0            0            0            0            0            1 
  0.35            0            0            0            0            0            0            0            0            0            0            0            0            0            0            0 
  0.40            0            0            0            0            0            0            0            0            0            0            0            0            0            0            0 
  0.45            0            0            0            0            0            0            0            0            0            0            0            0            0            0            0 
  0.50            0            0            0            0            0            0            0            0            0            0            0            0            0            0            0 
  }\plotdata;
		  \begin{axis} [
      axis lines=left,
      xlabel={$\alpha$},
      ylabel={samples},
      ymin=0, ymax=100,
      xmin=0.1, xmax=0.5,
      minor x tick num=3,
      minor y tick num=1,
      legend cell align=left,
      legend style={ draw=none, at={(1.35,1.0)}, },
      ybar interval=0.7,
    ]
    \addplot[red,fill=red]       table[ x index=0, y index=9,  ]                {\plotdata};
    \addplot[orange,fill=orange] table[ x index=0, y index=10, ]                {\plotdata};
    \addplot[yellow,fill=yellow] table[ x index=0, y index=11, ]                {\plotdata};
    \addplot[lime,fill=lime]     table[ x index=0, y index=12, ]                {\plotdata};
    \addplot[cyan,fill=cyan]     table[ x index=0, y index=13, ]                {\plotdata};
    \addplot[blue,fill=blue]     table[ x index=0, y index=14, ]                {\plotdata};
    \addplot[violet,fill=violet] table[ x index=0, y index=15, ]                {\plotdata};
    \legend{$L_c = 10$,$L_c = 11$,$L_c = 12$,$L_c = 13$,$L_c = 14$,$L_c = 15$,$L_c = 16$}
  \end{axis}
	    \end{tikzpicture}
	}
	\caption{ $R=\infty$ } \label{fig:NSTSM99}
    \end{subfigure}
     \caption{ The number of type 1 cases out of 100 cases that the D-Wave was unable to solve within 8192 samples. }
    \label{fig:NSTSM}
\end{figure}

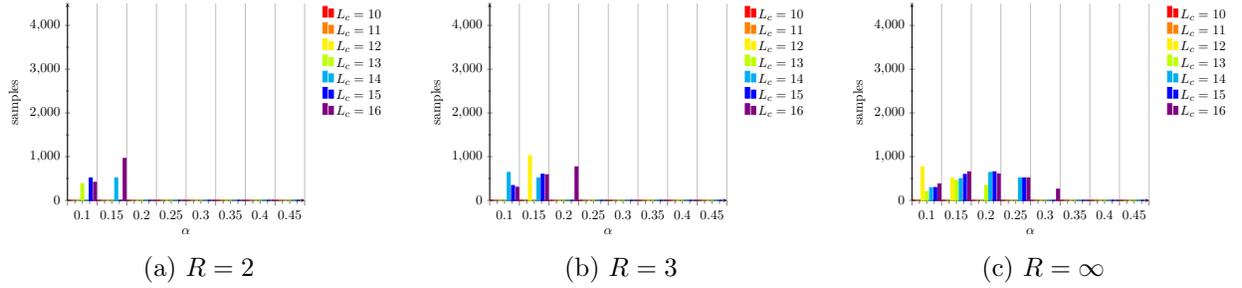
\begin{figure}
    \begin{subfigure}{0.32\textwidth}
	\scalebox{0.46}{
	    \begin{tikzpicture}
		  \pgfplotstableread{
 Alpha       mqcL02       mqcL03       mqcL04       mqcL05       mqcL06       mqcL07       mqcL08       mqcL09       mqcL10       mqcL11       mqcL12       mqcL13       mqcL14       mqcL15       mqcL16 
  0.05            0            0            0            0            0            0            0            0            0            0            0            0            0            0            0 
  0.10            0            0            0            0            0            0            0            0            0            0            0          384            0          512          413 
  0.15            0            0            0            0            0            0            0            0            0            0            0            0          512            0          960 
  0.20            0            0            0            0            0            0            0            0            0            0            0            0            0            0            0 
  0.25            0            0            0            0            0            0            0            0            0            0            0            0            0            0            0 
  0.30            0            0            0            0            0            0            0            0            0            0            0            0            0            0            0 
  0.35            0            0            0            0            0            0            0            0            0            0            0            0            0            0            0 
  0.40            0            0            0            0            0            0            0            0            0            0            0            0            0            0            0 
  0.45            0            0            0            0            0            0            0            0            0            0            0            0            0            0            0 
  0.50            0            0            0            0            0            0            0            0            0            0            0            0            0            0            0 
  }\plotdata;
		  \begin{axis} [
      axis lines=left,
      xlabel={$\alpha$},
      ylabel={samples},
      ymin=0, ymax=4500,
      xmin=0.1, xmax=0.5,
      minor x tick num=3,
      minor y tick num=1,
      legend cell align=left,
      legend style={ draw=none, at={(1.35,1.0)}, },
      ybar interval=0.7,
    ]
    \addplot[red,fill=red]       table[ x index=0, y index=9,  ]                {\plotdata};
    \addplot[orange,fill=orange] table[ x index=0, y index=10, ]                {\plotdata};
    \addplot[yellow,fill=yellow] table[ x index=0, y index=11, ]                {\plotdata};
    \addplot[lime,fill=lime]     table[ x index=0, y index=12, ]                {\plotdata};
    \addplot[cyan,fill=cyan]     table[ x index=0, y index=13, ]                {\plotdata};
    \addplot[blue,fill=blue]     table[ x index=0, y index=14, ]                {\plotdata};
    \addplot[violet,fill=violet] table[ x index=0, y index=15, ]                {\plotdata};
    \legend{$L_c = 10$,$L_c = 11$,$L_c = 12$,$L_c = 13$,$L_c = 14$,$L_c = 15$,$L_c = 16$}
  \end{axis}
	    \end{tikzpicture}
	}
	\caption{ $R=2$ } \label{fig:NASTSM2}
    \end{subfigure}
    \hspace*{\fill} 
    \begin{subfigure}{0.32\textwidth}
	\scalebox{0.46}{
	    \begin{tikzpicture}
		  \pgfplotstableread{
 Alpha       mqcL02       mqcL03       mqcL04       mqcL05       mqcL06       mqcL07       mqcL08       mqcL09       mqcL10       mqcL11       mqcL12       mqcL13       mqcL14       mqcL15       mqcL16 
  0.05            0            0            0            0            0            0            0            0            0            0            0            0            0            0            0 
  0.10            0            0            0            0            0            0            0            0            0            0            0            0          640          339          300 
  0.15            0            0            0            0            0            0            0            0            0            0         1024            0          512          597          587 
  0.20            0            0            0            0            0            0            0            0            0            0            0            0            0            0          768 
  0.25            0            0            0            0            0            0            0            0            0            0            0            0            0            0            0 
  0.30            0            0            0            0            0            0            0            0            0            0            0            0            0            0            0 
  0.35            0            0            0            0            0            0            0            0            0            0            0            0            0            0            0 
  0.40            0            0            0            0            0            0            0            0            0            0            0            0            0            0            0 
  0.45            0            0            0            0            0            0            0            0            0            0            0            0            0            0            0 
  0.50            0            0            0            0            0            0            0            0            0            0            0            0            0            0            0 
  }\plotdata;
		  \begin{axis} [
      axis lines=left,
      xlabel={$\alpha$},
      ylabel={samples},
      ymin=0, ymax=4500,
      xmin=0.1, xmax=0.5,
      minor x tick num=3,
      minor y tick num=1,
      legend cell align=left,
      legend style={ draw=none, at={(1.35,1.0)}, },
      ybar interval=0.7,
    ]
    \addplot[red,fill=red]       table[ x index=0, y index=9,  ]                {\plotdata};
    \addplot[orange,fill=orange] table[ x index=0, y index=10, ]                {\plotdata};
    \addplot[yellow,fill=yellow] table[ x index=0, y index=11, ]                {\plotdata};
    \addplot[lime,fill=lime]     table[ x index=0, y index=12, ]                {\plotdata};
    \addplot[cyan,fill=cyan]     table[ x index=0, y index=13, ]                {\plotdata};
    \addplot[blue,fill=blue]     table[ x index=0, y index=14, ]                {\plotdata};
    \addplot[violet,fill=violet] table[ x index=0, y index=15, ]                {\plotdata};
    \legend{$L_c = 10$,$L_c = 11$,$L_c = 12$,$L_c = 13$,$L_c = 14$,$L_c = 15$,$L_c = 16$}
  \end{axis}
	    \end{tikzpicture}
	}
	\caption{ $R=3$ } \label{fig:NASTSM3}
    \end{subfigure}
    \hspace*{\fill} 
    \begin{subfigure}{0.32\textwidth}
	\scalebox{0.46}{
	    \begin{tikzpicture}
		  \pgfplotstableread{
 Alpha       mqcL02       mqcL03       mqcL04       mqcL05       mqcL06       mqcL07       mqcL08       mqcL09       mqcL10       mqcL11       mqcL12       mqcL13       mqcL14       mqcL15       mqcL16 
  0.05            0            0            0            0            0            0            0            0            0            0            0            0            0            0            0 
  0.10            0            0            0            0            0            0            0            0            0            0          768          204          290          295          377 
  0.15            0            0            0            0            0            0            0            0            0            0          512          457          498          594          653 
  0.20            0            0            0            0            0            0            0            0            0            0            0          341          640          654          604 
  0.25            0            0            0            0            0            0            0            0            0            0            0            0          512          512          512 
  0.30            0            0            0            0            0            0            0            0            0            0            0            0            0            0          256 
  0.35            0            0            0            0            0            0            0            0            0            0            0            0            0            0            0 
  0.40            0            0            0            0            0            0            0            0            0            0            0            0            0            0            0 
  0.45            0            0            0            0            0            0            0            0            0            0            0            0            0            0            0 
  0.50            0            0            0            0            0            0            0            0            0            0            0            0            0            0            0 
  }\plotdata;
		  \begin{axis} [
      axis lines=left,
      xlabel={$\alpha$},
      ylabel={samples},
      ymin=0, ymax=4500,
      xmin=0.1, xmax=0.5,
      minor x tick num=3,
      minor y tick num=1,
      legend cell align=left,
      legend style={ draw=none, at={(1.35,1.0)}, },
      ybar interval=0.7,
    ]
    \addplot[red,fill=red]       table[ x index=0, y index=9,  ]                {\plotdata};
    \addplot[orange,fill=orange] table[ x index=0, y index=10, ]                {\plotdata};
    \addplot[yellow,fill=yellow] table[ x index=0, y index=11, ]                {\plotdata};
    \addplot[lime,fill=lime]     table[ x index=0, y index=12, ]                {\plotdata};
    \addplot[cyan,fill=cyan]     table[ x index=0, y index=13, ]                {\plotdata};
    \addplot[blue,fill=blue]     table[ x index=0, y index=14, ]                {\plotdata};
    \addplot[violet,fill=violet] table[ x index=0, y index=15, ]                {\plotdata};
    \legend{$L_c = 10$,$L_c = 11$,$L_c = 12$,$L_c = 13$,$L_c = 14$,$L_c = 15$,$L_c = 16$}
  \end{axis}
	    \end{tikzpicture}
	}
	\caption{ $R=\infty$ } \label{fig:NASTSM99}
    \end{subfigure}
     \caption{ The average number of samples needed by MQC to solve a type 1 FCL that the D-Wave was 
     unable to solve.  }
    \label{fig:NASTSM}
\end{figure}

\begin{figure}
    \begin{subfigure}{0.32\textwidth}
	\scalebox{0.48}{
	    \begin{tikzpicture}
		  \pgfplotstableread{
 Alpha       rawL02       rawL03       rawL04       rawL05       rawL06       rawL07       rawL08       rawL09       rawL10       rawL11       rawL12       rawL13       rawL14       rawL15       rawL16 
  0.05            0            0            0            0            0            0            0            0            0            0            0            0            0            0            0 
  0.10            0            0            0            0            0            0            0            0            0            0            0            1            1            0            1 
  0.15            0            0            0            0            0            0            0            0            0            0            0            0            2            5            5 
  0.20            0            0            0            0            0            0            0            0            0            0            0            0            0            1            1 
  0.25            0            0            0            0            0            0            0            0            0            0            0            0            0            0            0 
  0.30            0            0            0            0            0            0            0            0            0            0            0            0            0            0            0 
  0.35            0            0            0            0            0            0            0            0            0            0            0            0            0            0            0 
  0.40            0            0            0            0            0            0            0            0            0            0            0            0            0            0            0 
  0.45            0            0            0            0            0            0            0            0            0            0            0            0            0            0            0 
  0.50            0            0            0            0            0            0            0            0            0            0            0            0            0            0            0 
  }\plotdata;
		  \begin{axis} [
      axis lines=left,
      xlabel={$\alpha$},
      ylabel={samples},
      ymin=0, ymax=100,
      xmin=0.1, xmax=0.5,
      minor x tick num=3,
      minor y tick num=1,
      legend cell align=left,
      legend style={ draw=none, at={(1.35,1.0)}, },
      ybar interval=0.7,
    ]
    \addplot[red,fill=red]       table[ x index=0, y index=9,  ]                {\plotdata};
    \addplot[orange,fill=orange] table[ x index=0, y index=10, ]                {\plotdata};
    \addplot[yellow,fill=yellow] table[ x index=0, y index=11, ]                {\plotdata};
    \addplot[lime,fill=lime]     table[ x index=0, y index=12, ]                {\plotdata};
    \addplot[cyan,fill=cyan]     table[ x index=0, y index=13, ]                {\plotdata};
    \addplot[blue,fill=blue]     table[ x index=0, y index=14, ]                {\plotdata};
    \addplot[violet,fill=violet] table[ x index=0, y index=15, ]                {\plotdata};
    \legend{$L_c = 10$,$L_c = 11$,$L_c = 12$,$L_c = 13$,$L_c = 14$,$L_c = 15$,$L_c = 16$}
  \end{axis}
	    \end{tikzpicture}
	}
	\caption{ $R=2$ } \label{fig:NSTSC2}
    \end{subfigure}
    \hspace*{\fill} 
    \begin{subfigure}{0.32\textwidth}
	\scalebox{0.48}{
	    \begin{tikzpicture}
		  \pgfplotstableread{
 Alpha       rawL02       rawL03       rawL04       rawL05       rawL06       rawL07       rawL08       rawL09       rawL10       rawL11       rawL12       rawL13       rawL14       rawL15       rawL16 
  0.05            0            0            0            0            0            0            0            0            0            0            0            0            0            0            0 
  0.10            0            0            0            0            0            0            0            0            0            0            0            1            1            6            7 
  0.15            0            0            0            0            0            0            0            0            0            0            0            4           15           20           22 
  0.20            0            0            0            0            0            0            0            0            0            0            0            1            8            5            0 
  0.25            0            0            0            0            0            0            0            0            0            0            0            0            0            1            0 
  0.30            0            0            0            0            0            0            0            0            0            0            0            0            0            0            0 
  0.35            0            0            0            0            0            0            0            0            0            0            0            0            0            0            0 
  0.40            0            0            0            0            0            0            0            0            0            0            0            0            0            0            0 
  0.45            0            0            0            0            0            0            0            0            0            0            0            0            0            0            0 
  0.50            0            0            0            0            0            0            0            0            0            0            0            0            0            0            0 
  }\plotdata;
		  \begin{axis} [
      axis lines=left,
      xlabel={$\alpha$},
      ylabel={samples},
      ymin=0, ymax=100,
      xmin=0.1, xmax=0.5,
      minor x tick num=3,
      minor y tick num=1,
      legend cell align=left,
      legend style={ draw=none, at={(1.35,1.0)}, },
      ybar interval=0.7,
    ]
    \addplot[red,fill=red]       table[ x index=0, y index=9,  ]                {\plotdata};
    \addplot[orange,fill=orange] table[ x index=0, y index=10, ]                {\plotdata};
    \addplot[yellow,fill=yellow] table[ x index=0, y index=11, ]                {\plotdata};
    \addplot[lime,fill=lime]     table[ x index=0, y index=12, ]                {\plotdata};
    \addplot[cyan,fill=cyan]     table[ x index=0, y index=13, ]                {\plotdata};
    \addplot[blue,fill=blue]     table[ x index=0, y index=14, ]                {\plotdata};
    \addplot[violet,fill=violet] table[ x index=0, y index=15, ]                {\plotdata};
    \legend{$L_c = 10$,$L_c = 11$,$L_c = 12$,$L_c = 13$,$L_c = 14$,$L_c = 15$,$L_c = 16$}
  \end{axis}
	    \end{tikzpicture}
	}
	\caption{ $R=3$ } \label{fig:NSTSC3}
    \end{subfigure}
    \hspace*{\fill} 
    \begin{subfigure}{0.32\textwidth}
	\scalebox{0.48}{
	    \begin{tikzpicture}
		  \pgfplotstableread{
 Alpha       rawL02       rawL03       rawL04       rawL05       rawL06       rawL07       rawL08       rawL09       rawL10       rawL11       rawL12       rawL13       rawL14       rawL15       rawL16
  0.05            0            0            0            0            0            0            0            0            0            0            0            0            0            0            0 
  0.10            0            0            0            0            0            0            0            0            0            0            1            1            5           16           31 
  0.15            0            0            0            0            0            0            0            0            1            2            6           20           52           77           89 
  0.20            0            0            0            0            0            0            0            0            0            1            8           22           49           76           66 
  0.25            0            0            0            0            0            0            0            0            0            0            1           11           18           44           24 
  0.30            0            0            0            0            0            0            0            0            0            0            0            1            8           10            0 
  0.35            0            0            0            0            0            0            0            0            0            0            0            0            1            1            6 
  0.40            0            0            0            0            0            0            0            0            0            0            0            0            0            0            0 
  0.45            0            0            0            0            0            0            0            0            0            0            0            0            0            0            1 
  0.50            0            0            0            0            0            0            0            0            0            0            0            0            0            0            0 
  }\plotdata;
		  \begin{axis} [
      axis lines=left,
      xlabel={$\alpha$},
      ylabel={samples},
      ymin=0, ymax=100,
      xmin=0.1, xmax=0.5,
      minor x tick num=3,
      minor y tick num=1,
      legend cell align=left,
      legend style={ draw=none, at={(1.35,1.0)}, },
      ybar interval=0.7,
    ]
    \addplot[red,fill=red]       table[ x index=0, y index=9,  ]                {\plotdata};
    \addplot[orange,fill=orange] table[ x index=0, y index=10, ]                {\plotdata};
    \addplot[yellow,fill=yellow] table[ x index=0, y index=11, ]                {\plotdata};
    \addplot[lime,fill=lime]     table[ x index=0, y index=12, ]                {\plotdata};
    \addplot[cyan,fill=cyan]     table[ x index=0, y index=13, ]                {\plotdata};
    \addplot[blue,fill=blue]     table[ x index=0, y index=14, ]                {\plotdata};
    \addplot[violet,fill=violet] table[ x index=0, y index=15, ]                {\plotdata};
    \legend{$L_c = 10$,$L_c = 11$,$L_c = 12$,$L_c = 13$,$L_c = 14$,$L_c = 15$,$L_c = 16$}
  \end{axis}
	    \end{tikzpicture}
	}
	\caption{ $R=\infty$ } \label{fig:NSTSC99}
    \end{subfigure}
     \caption{ The number of type 2 cases out of 100 cases that the D-Wave was unable to solve within 8192 samples. }
    \label{fig:NSTSC}
\end{figure}
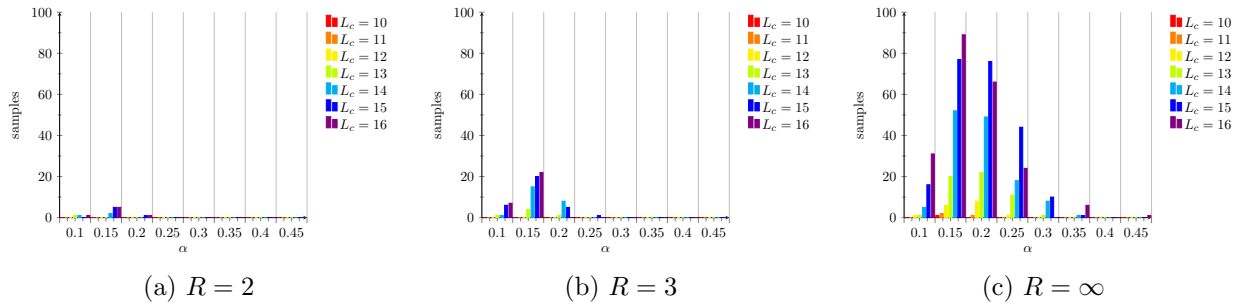

In some cases, the problems are so hard that the the D-Wave is unable to find the answer within 8192 samples.
Figure \ref{fig:NSTSM} (type 1) and figure \ref{fig:NSTSC} (type 2) show the number of cases 
out of 100 that D-Wave does not solve within 8192 samples. 
It appears to be worse for type 2 problems than for type 1 problems.
This only occurred for D-Wave partitions sizes corresponding to $c=10$ or greater.
In all such cases that were run, MQC post-processing found the answer.

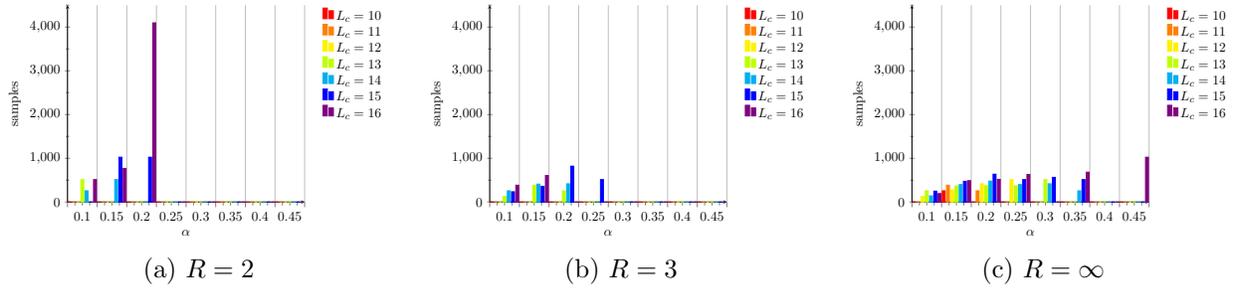
\begin{figure}
    \begin{subfigure}{0.32\textwidth}
	\scalebox{0.46}{
	    \begin{tikzpicture}
		  \pgfplotstableread{
 Alpha       mqcL02       mqcL03       mqcL04       mqcL05       mqcL06       mqcL07       mqcL08       mqcL09       mqcL10       mqcL11       mqcL12       mqcL13       mqcL14       mqcL15       mqcL16 
  0.05            0            0            0            0            0            0            0            0            0            0            0            0            0            0            0 
  0.10            0            0            0            0            0            0            0            0            0            0            0          512          256            0          512 
  0.15            0            0            0            0            0            0            0            0            0            0            0            0          512         1024          768 
  0.20            0            0            0            0            0            0            0            0            0            0            0            0            0         1024         4096 
  0.25            0            0            0            0            0            0            0            0            0            0            0            0            0            0            0 
  0.30            0            0            0            0            0            0            0            0            0            0            0            0            0            0            0 
  0.35            0            0            0            0            0            0            0            0            0            0            0            0            0            0            0 
  0.40            0            0            0            0            0            0            0            0            0            0            0            0            0            0            0 
  0.45            0            0            0            0            0            0            0            0            0            0            0            0            0            0            0 
  0.50            0            0            0            0            0            0            0            0            0            0            0            0            0            0            0 
  }\plotdata;
		  \begin{axis} [
      axis lines=left,
      xlabel={$\alpha$},
      ylabel={samples},
      ymin=0, ymax=4500,
      xmin=0.1, xmax=0.5,
      minor x tick num=3,
      minor y tick num=1,
      legend cell align=left,
      legend style={ draw=none, at={(1.35,1.0)}, },
      ybar interval=0.7,
    ]
    \addplot[red,fill=red]       table[ x index=0, y index=9,  ]                {\plotdata};
    \addplot[orange,fill=orange] table[ x index=0, y index=10, ]                {\plotdata};
    \addplot[yellow,fill=yellow] table[ x index=0, y index=11, ]                {\plotdata};
    \addplot[lime,fill=lime]     table[ x index=0, y index=12, ]                {\plotdata};
    \addplot[cyan,fill=cyan]     table[ x index=0, y index=13, ]                {\plotdata};
    \addplot[blue,fill=blue]     table[ x index=0, y index=14, ]                {\plotdata};
    \addplot[violet,fill=violet] table[ x index=0, y index=15, ]                {\plotdata};
    \legend{$L_c = 10$,$L_c = 11$,$L_c = 12$,$L_c = 13$,$L_c = 14$,$L_c = 15$,$L_c = 16$}
  \end{axis}
	    \end{tikzpicture}
	}
	\caption{ $R=2$ } \label{fig:NASTSC2}
    \end{subfigure}
    \hspace*{\fill} 
    \begin{subfigure}{0.32\textwidth}
	\scalebox{0.46}{
	    \begin{tikzpicture}
		  \pgfplotstableread{
 Alpha       mqcL02       mqcL03       mqcL04       mqcL05       mqcL06       mqcL07       mqcL08       mqcL09       mqcL10       mqcL11       mqcL12       mqcL13       mqcL14       mqcL15       mqcL16 
  0.05            0            0            0            0            0            0            0            0            0            0            0            0            0            0            0 
  0.10            0            0            0            0            0            0            0            0            0            0            0          128          256          234          384 
  0.15            0            0            0            0            0            0            0            0            0            0            0          384          409          358          605 
  0.20            0            0            0            0            0            0            0            0            0            0            0          256          416          819            0 
  0.25            0            0            0            0            0            0            0            0            0            0            0            0            0          512            0 
  0.30            0            0            0            0            0            0            0            0            0            0            0            0            0            0            0 
  0.35            0            0            0            0            0            0            0            0            0            0            0            0            0            0            0 
  0.40            0            0            0            0            0            0            0            0            0            0            0            0            0            0            0 
  0.45            0            0            0            0            0            0            0            0            0            0            0            0            0            0            0 
  0.50            0            0            0            0            0            0            0            0            0            0            0            0            0            0            0 
  }\plotdata;
		  \begin{axis} [
      axis lines=left,
      xlabel={$\alpha$},
      ylabel={samples},
      ymin=0, ymax=4500,
      xmin=0.1, xmax=0.5,
      minor x tick num=3,
      minor y tick num=1,
      legend cell align=left,
      legend style={ draw=none, at={(1.35,1.0)}, },
      ybar interval=0.7,
    ]
    \addplot[red,fill=red]       table[ x index=0, y index=9,  ]                {\plotdata};
    \addplot[orange,fill=orange] table[ x index=0, y index=10, ]                {\plotdata};
    \addplot[yellow,fill=yellow] table[ x index=0, y index=11, ]                {\plotdata};
    \addplot[lime,fill=lime]     table[ x index=0, y index=12, ]                {\plotdata};
    \addplot[cyan,fill=cyan]     table[ x index=0, y index=13, ]                {\plotdata};
    \addplot[blue,fill=blue]     table[ x index=0, y index=14, ]                {\plotdata};
    \addplot[violet,fill=violet] table[ x index=0, y index=15, ]                {\plotdata};
    \legend{$L_c = 10$,$L_c = 11$,$L_c = 12$,$L_c = 13$,$L_c = 14$,$L_c = 15$,$L_c = 16$}
  \end{axis}
	    \end{tikzpicture}
	}
	\caption{ $R=3$ } \label{fig:NASTSC3}
    \end{subfigure}
    \hspace*{\fill} 
    \begin{subfigure}{0.32\textwidth}
	\scalebox{0.46}{
	    \begin{tikzpicture}
		  \pgfplotstableread{
 Alpha       mqcL02       mqcL03       mqcL04       mqcL05       mqcL06       mqcL07       mqcL08       mqcL09       mqcL10       mqcL11       mqcL12       mqcL13       mqcL14       mqcL15       mqcL16 
  0.05            0            0            0            0            0            0            0            0            0            0            0            0            0            0            0 
  0.10            0            0            0            0            0            0            0            0            0            0          128          256          140          248          192 
  0.15            0            0            0            0            0            0            0            0          256          384          277          364          398          473          488 
  0.20            0            0            0            0            0            0            0            0            0          256          416          372          478          636          519 
  0.25            0            0            0            0            0            0            0            0            0            0          512          372          398          512          629 
  0.30            0            0            0            0            0            0            0            0            0            0            0          512          416          563            0 
  0.35            0            0            0            0            0            0            0            0            0            0            0            0          256          512          682 
  0.40            0            0            0            0            0            0            0            0            0            0            0            0            0            0            0 
  0.45            0            0            0            0            0            0            0            0            0            0            0            0            0            0         1024 
  0.50            0            0            0            0            0            0            0            0            0            0            0            0            0            0            0 
  }\plotdata;
		  \begin{axis} [
      axis lines=left,
      xlabel={$\alpha$},
      ylabel={samples},
      ymin=0, ymax=4500,
      xmin=0.1, xmax=0.5,
      minor x tick num=3,
      minor y tick num=1,
      legend cell align=left,
      legend style={ draw=none, at={(1.35,1.0)}, },
      ybar interval=0.7,
    ]
    \addplot[red,fill=red]       table[ x index=0, y index=9,  ]                {\plotdata};
    \addplot[orange,fill=orange] table[ x index=0, y index=10, ]                {\plotdata};
    \addplot[yellow,fill=yellow] table[ x index=0, y index=11, ]                {\plotdata};
    \addplot[lime,fill=lime]     table[ x index=0, y index=12, ]                {\plotdata};
    \addplot[cyan,fill=cyan]     table[ x index=0, y index=13, ]                {\plotdata};
    \addplot[blue,fill=blue]     table[ x index=0, y index=14, ]                {\plotdata};
    \addplot[violet,fill=violet] table[ x index=0, y index=15, ]                {\plotdata};
    \legend{$L_c = 10$,$L_c = 11$,$L_c = 12$,$L_c = 13$,$L_c = 14$,$L_c = 15$,$L_c = 16$}
  \end{axis}
	    \end{tikzpicture}
	}
	\caption{ $R=\infty$ } \label{fig:NASTSC99}
    \end{subfigure}
     \caption{  The average number of samples needed by MQC to solve a type 2 FCL that the D-Wave was 
     unable to solve.  }
    \label{fig:NASTSC}
\end{figure}

Figures \ref{fig:NASTSM} (type 1) and \ref{fig:NASTSC} (type 2) show the number of cases needed
on average for MQC post-processing to find a solution that the D-Wave alone did not solve with 
8192 samples.
Note that, in all the cases that the D-Wave was unable to solve in 8192 samples, 
the D-Wave with MQC post-processing was able to solve, and in most case in less than 1000 samples.

\section{Conclusion}\label{sec:Conclusion}

The FCL problem class has been used to test the relative performance of the D-Wave with and without
MQC post-processing.
FCL problems were generated with a wide variety of difficulty, from problems that the D-Wave always solved to
problems that it never solved.
MQC post-processing has shown exceptional performance over the D-Wave without post-processing.
MQC post-processing provded little benefit if the problem was easy, as one would expect.
However, MQC post-processing has demostrated a 10 ot 20 times improvement in execution speed for problems of moderate
difficulty (problems that the D-Wave can solve) and has solved problems the D-Wave was unable to 
solve alone even though 81920 samples were requested.
These unsolved problems were often solved by MQC post-processing with less than D-Wave 1000 samples.
The D-Wave with MQC post-processing was able to solve all cases of FCL, that were attempted.

\section*{Acknowledgment}

The author would like to thank Michael Little and Marjorie Cole of the NASA Advanced Information Systems 
Technology Office for their continued support for this research effort under grant NNH16ZDA001N-AIST16-0091 
and to the NASA Ames Research Center for providing access to the D-Wave quantum annealing computer. 
In addition, the author thanks the NSF funded Center for Hybrid Multicore Productivity Research, 
D-Wave Systems for their support and access to their computational resources, and  
for quantum computing system resources supported by the Quantum Computing Institute at Oak Ridge National Laboratory.

\bibliography{QuantumBib}{}

\begin{thebibliography}{8}
\providecommand{\natexlab}[1]{#1}
\providecommand{\url}[1]{\texttt{#1}}
\expandafter\ifx\csname urlstyle\endcsname\relax
  \providecommand{\doi}[1]{doi: #1}\else
  \providecommand{\doi}{doi: \begingroup \urlstyle{rm}\Url}\fi

\bibitem[{Dorband}(2017)]{Dorband17}
J.~E. {Dorband}.
\newblock {Improving the Accuracy of an Adiabatic Quantum Computer}.
\newblock \emph{ArXiv e-prints}, May 2017.

\bibitem[{Farhi} et~al.(2000){Farhi}, {Goldstone}, {Gutmann}, and
  {Sipser}]{Farhi00}
E.~{Farhi}, J.~{Goldstone}, S.~{Gutmann}, and M.~{Sipser}.
\newblock {Quantum Computation by Adiabatic Evolution}.
\newblock \emph{eprint arXiv:quant-ph/0001106}, January 2000.

\bibitem[{Hen} et~al.(2015){Hen}, {Job}, {Albash}, {R{\o}nnow}, {Troyer}, and
  {Lidar}]{Hen2015}
I.~{Hen}, J.~{Job}, T.~{Albash}, T.~F. {R{\o}nnow}, M.~{Troyer}, and D.~A.
  {Lidar}.
\newblock {Probing for quantum speedup in spin-glass problems with planted
  solutions}.
\newblock 92\penalty0 (4):\penalty0 042325, October 2015.
\newblock \doi{10.1103/PhysRevA.92.042325}.

\bibitem[Inc.(2013)]{Dwave13}
D-Wave~Systems Inc.
\newblock {The D-Wave 2X Quantum Computer: Technology Overview}.
\newblock \url{http://www.dwavesys.com/resources/publications}, 2013.
\newblock [Online; accessed 18-May-2016].

\bibitem[{King} et~al.(2015){King}, {Lanting}, and {Harris}]{King2015}
A.~D. {King}, T.~{Lanting}, and R.~{Harris}.
\newblock {Performance of a quantum annealer on range-limited constraint
  satisfaction problems}.
\newblock \emph{ArXiv e-prints}, February 2015.

\bibitem[{King} et~al.(2017){King}, {Yarkoni}, {Raymond}, {Ozfidan}, {King},
  {Nevisi}, {Hilton}, and {McGeoch}]{King2017}
J.~{King}, S.~{Yarkoni}, J.~{Raymond}, I.~{Ozfidan}, A.~D. {King}, M.~M.
  {Nevisi}, J.~P. {Hilton}, and C.~C. {McGeoch}.
\newblock {Quantum Annealing amid Local Ruggedness and Global Frustration}.
\newblock \emph{ArXiv e-prints}, January 2017.

\bibitem[{Mandr{\`a}} and {Katzgraber}(2018)]{Mandra18}
S.~{Mandr{\`a}} and H.~G. {Katzgraber}.
\newblock {A deceptive step towards quantum speedup detection}.
\newblock \emph{Quantum Science and Technology}, 3\penalty0 (4):\penalty0
  04LT01, October 2018.
\newblock \doi{10.1088/2058-9565/aac8b2}.

\bibitem[{Santoro} and {Tosatti}(2008)]{Giuseppe08}
Giuseppe {Santoro} and Erio {Tosatti}.
\newblock Optimization using quantum mechanics: quantum annealing through
  adiabatic evolution.
\newblock \emph{Journal of Physics A: Mathematical and Theoretical},
  41\penalty0 (20):\penalty0 209801, 2008.
\newblock URL \url{http://stacks.iop.org/1751-8121/41/i=20/a=209801}.

\end{thebibliography}

\end{document}